\documentclass[11pt,final,onecolumn,twoside]{IEEEtran}

\usepackage{amsmath}
\usepackage{amssymb}
\usepackage{graphicx}
\usepackage{bm}
\usepackage{cite}

\newtheorem{theorem}{Theorem}
\newtheorem{lemma}[theorem]{Lemma}

\begin{document}

\title{State Estimation Over Wireless Channels Using Multiple Sensors: Asymptotic Behaviour and Optimal Power Allocation}

\author{Alex S. Leong, Subhrakanti Dey, and Jamie S. Evans
  \thanks{The authors are with the ARC Special Research Center for Ultra-Broadband Information Networks (CUBIN, an affiliated program of National ICT Australia),
  Department of Electrical and Electronic Engineering,
  University of Melbourne, Parkville, Vic. 3010, Australia.}
  \thanks{Tel: 613-8344-3819. Fax: 613-8344-6678. E-mail \tt\{asleong, sdey, jse\}@unimelb.edu.au}
  \thanks{This work was supported by the Australian Research Council}
  }
\maketitle
\thispagestyle{empty}
\begin{abstract}
This paper considers state estimation of linear systems using analog amplify and forwarding with multiple sensors, for both multiple access and orthogonal access schemes. Optimal state estimation  can be achieved at the fusion center using a time varying Kalman filter. We show that in many situations, the estimation error covariance decays at a rate of $1/M$ when the number of sensors $M$ is large. We consider optimal allocation of transmission powers that 1) minimizes the sum power usage subject to an error covariance constraint and 2) minimizes the error covariance subject to a sum power constraint. In the case of fading channels with channel state information the optimization problems are solved using a greedy approach, while for fading channels without channel state information but with channel statistics available a sub-optimal linear estimator is derived. 
\end{abstract}

\begin{keywords}
Distributed estimation, Kalman filtering, power allocation, scaling laws, sensor networks
\end{keywords}

\section{Introduction}
Wireless sensor networks are collections of sensors which can communicate with each other or to a central node or base station through wireless links. Potential uses include environment and infrastructure monitoring, healthcare 
and military applications, to name a few. 
Often these sensors will have limited energy and computational ability which imposes severe constraints on system design, and signal processing algorithms which can efficiently utilise these resources have attracted great interest. 

In recent years there has been a considerable literature on estimation and detection schemes  designed specifically for use in wireless sensor networks. 
Work on detection in wireless sensor networks include \cite{ChamberlandVeeravalli} which studies the asymptotic optimality of using identical sensors in the presence of energy constraints, and \cite{RagoWillettBarShalom,ChenJiangKasetkasemVarshney,JiangChen} which derives fusion rules for distributed detection in the presence of fading. Parameter estimation or estimation of constant signals is studied in e.g. \cite{RibeiroGiannakis1,Xiao_TSP,SchizasGiannakisLuo,WuHuangLee} where issues of quantization and optimization of power usage are addressed.
Type based methods for detection and estimation of discrete sources are proposed and analyzed in \cite{MergenTong,MergenNawareTong,LiuSayeed}. Estimation of fields is considered has been considered in e.g. \cite{NowakMitraWillett,WangIshwarSaligrama,ZhangMouraKrogh}. 

A promising scheme for distributed estimation in sensor networks is analog amplify and forward \cite{GastparVetterli} (in distributed detection analog forwarding has also been considered in e.g. \cite{LiDai,ZhangPoorChiang}), where measurements from the sensors are transmitted directly (possibly scaled) to the fusion center without any coding, which is motivated by optimality results on uncoded transmissions in point-to-point links \cite{Goblick,GastparRimoldiVetterli}. (Other related information theoretic results include \cite{ViswanathanBerger,Ishwar_JSAC}.) Analog forwarding schemes are attractive due to their simplicity as well as the possibility of real-time processing since there is no coding delay.
In \cite{GastparVetterli} the asymptotic (large number of sensors) optimality of analog forwarding for estimating an i.i.d. scalar Gaussian process was shown, and exact optimality was later proved for a ``symmetric'' sensor network \cite{Gastpar_optimality}. Analog forwarding with optimal power allocation is studied in \cite{Xiao_coherent_TSP} and \cite{Cui_TSP} for multi-access and orthogonal schemes respectively. Lower bounds and asymptotic optimality results for estimating independent vector processes, is addressed in \cite{Gastpar_JSAC}. Estimation with correlated data between sensors is studied in  \cite{BahceciKhandani,FangLi}. Other aspects of the analog forwading technique that have been studied include the use of different network topologies \cite{ThatteMitra}, other multiple access schemes such as slotted ALOHA \cite{HongLeiChi}, and consideration of the impact of channel estimation errors \cite{Senol} on estimation performance. 

Most of the previous work on analog forwarding have dealt with estimation of processes which are either constant or i.i.d over time. In this paper we will address the estimation of \emph{dynamical systems} using analog forwarding of measurements. In particular, we will consider the problem of state estimation of discrete-time linear systems using multiple sensors. As is well known, optimal state estimation of a linear system can be achieved using a Kalman filter. Other work on Kalman filtering in sensor networks include  studies of optimal sensor data quantization \cite{ZhangLi},  Kalman filtering using one bit quantized observations \cite{Ribeiro} where performance is shown to lie within a constant factor of the standard Kalman filter, and estimation of random fields with reduced order Kalman filters \cite{ZhangMouraKrogh}. 
Another related area with a rich history is that of distributed Kalman filtering, where the main objectives include doing local processing at the individual sensor level to reduce the computations required at the fusion center \cite{WillnerChangDunn,HashemipourRoyLaub}, or to form estimates at each of the individual sensors in a completely decentralized fashion without any fusion center \cite{RaoDurrantWhyte}. However in our work we assume that computational resources available at the sensors are limited so that they will only take measurements and then transmit them to the fusion center for further processing, using uncoded analog forwarding. 

{\em Summary of Contributions}:\\
In this paper we will mainly focus on estimation of scalar\footnote{By scalar linear system we mean that both the states and individual sensor measurements are scalar.} linear dynamical systems using multiple sensors, as the vector case introduces additional difficulties such that only partial results can be obtained. We will be interested in deriving the asymptotic behaviour of the error covariance with respect to the number of sensors for these schemes, as well 
as optimal transmission power allocation to the sensors under a constraint on the error covariance at the fusion center, or a sum power constraint at the sensor transmitters. We consider both static and fading channels, and in the context 
of fading channels, we consider various levels of availability of channel state information (CSI) at the transmitters 
and the fusion center. More specifically, we make the following key contributions:
\begin{itemize}
\item We show that (for static channels with full CSI) for the multi-access scheme, the asymptotic estimation error covariance can be driven to the process noise covariance (which is the minimum attainable error) as the number of sensors $M$ goes to infinity, even when the transmitted signals from each sensor is scaled by $\frac{1}{\sqrt{M}}$ (which implies that total 
transmission power across all sensors remains bounded while each sensor's transmission power goes to zero). 
This is a particularly attractive result since sensor networks operate in a energy limited environment.
For the orthogonal access scheme, this result holds when the transmitted signals are unscaled, but does not hold when the transmitted signals are scaled by $\frac{1}{\sqrt{M}}$. 
\item The convergence rate of these asymptotic results (when they hold) is shown to be $\frac{1}{M}$, although it is seen via simulation results that the asymptotic approximations 
are quite accurate even for $M=20$ to $30$ sensors.  

\item In the case of a small to moderate number of sensors, we derive a comprehensive set of optimal sensor transmit power allocation schemes for multi-access and orthogonal medium access schemes over both static and fading channels.  For static channels, we minimize total transmission power at the sensors subject to a constraint on the 
steady state Kalman estimation error covariance, and also solve a corresponding converse problem: minimizing steady state error covariance subject to a sum power constraint at the sensor transmitters. For fading channels (with full CSI), 
we solve similar optimization problems, except that the error covariance (either in the objective 
function or the constraint) is considered at a per time instant basis, since there is no well defined steady state error covariance in this case. 
For the fading channel case with no CSI (either amplitude or phase), the results are derived for the best linear estimator which relies on channel statistics information  and can be applied to non-zero mean fading channels. 
It is shown that 
these optimization problems can be posed as convex optimization problems.
Moreover, the optimization problems will turn out to be very similar to problems previously studied in the literature (albeit in the context of distributed estimation of a static random source), namely \cite{Xiao_coherent_TSP,Cui_TSP}, and can actually be solved in closed form. 

\item Numerical results demonstrate that for static channels, optimal power allocation results in more benefit for the orthogonal medium access scheme 
compared to the multi-access scheme, whereas for fading channels, it is seen that having full CSI is clearly beneficial for both schemes, although the performance 
improvement via the optimal power allocation scheme is more substantial for the orthogonal scheme than the multi-access scheme.    
\end{itemize}

The rest of the paper 
is organized as follows.  
Section \ref{model_sec} specifies our scalar models and preliminaries, and gives a number of examples between multi-access and orthogonal access schemes, which show that in general one scheme does not always perform better than the other. We investigate the asymptotic behaviour for a large number of sensors $M$ in Section \ref{asymptotic_sec}. Power allocation is considered in Section \ref{optim_sec}, where we formulate and solve optimization problems for 1) an error covariance constraint and 2) a sum power constraint. We first do this for static channels, before focusing on fading channels. In the case where we have channel state information (CSI) we use a greedy approach by performing the optimization at each time step. When we don't have CSI, we will derive a sub-optimal linear estimator similar to \cite{Nahi,Rajasekaran,HadidiSchwartz}, which can be used for non-zero mean fading. Numerical studies are presented in Section \ref{numerical_sec}. Extensions of our model to vector and MIMO systems is considered in Section \ref{extensions_sec}, where we formulate the models and optimization problems, and outline some of the difficulties involved. 

\section{Models and preliminaries}
\label{model_sec}
Throughout this paper, $i$ represents the sensor index and $k$ represents the time index.
Let the scalar linear system be
$$ x_{k+1} = a x_k + w_k$$
with the $M$ sensors each observing
$$ y_{i,k} = c_i x_k + v_{i,k} , i = 1, \dots , M$$
with $w_k$ and $v_{i,k}$ being zero-mean Gaussians having variances $\sigma_w^2$ and $\sigma_i^2$ respectively, with the $v_{i,k}$'s being independent between sensors. Note that the sensors can have different observation matrices $c_i$ and measurement noise variances $\sigma_i^2$, and
we allow $a$ and $c_i$ to take on both positive and negative values.
It is assumed that the parameters $a, c_i, \sigma_w^2$ and $\sigma_i^2$ are known.\footnote{We assume that these parameters are static or very slowly time-varying, and hence can be accurately determined beforehand using appropriate parameter estimation/system identification algorithms.} Furthermore, we assume that the system is stable, i.e. $|a| < 1$. 

\subsection{Multi-access scheme}
In the (non-orthogonal) multi-access scheme the fusion center receives the sum
\begin{equation}
\label{mac_complex_received}
\tilde{z}_k = \sum_{i=1}^M \tilde{\alpha}_{i,k} \tilde{h}_{i,k} y_{i,k} + \tilde{n}_k
\end{equation}
where $\tilde{n}_k$ is zero-mean complex Gaussian with variance $2\sigma_n^2$, $\tilde{h}_{i,k}$ are the complex-valued 
channel gains, and $\tilde{\alpha}_{i,k}$ are the complex-valued multiplicative amplification factors in an amplify and forward scheme. 
We assume that all transmitters have access to their complex channel state information (CSI),\footnote{The case where the channel gains are unknown but channel statistics are available is addressed in Section \ref{no_CSI_sec}. This can also be used to model the situation where perfect phase synchronization cannot be achieved \cite{Gastpar_JSAC}.}  and the amplification factors have the form
$$\tilde{\alpha}_{i,k} = \alpha_{i,k} \frac{\tilde{h}_{i,k}^*}{|\tilde{h}_{i,k}|}$$
where $\alpha_{i,k}$ is real-valued, i.e. we assume distributed transmitter beamforming.  Defining $h_{i,k} \equiv |\tilde{h}_{i,k}|$, $z_k \equiv \Re[\tilde{z}_k]$, $n_k \equiv \Re[\tilde{n}_k]$, we then have 
\begin{equation}
\label{coherent_sum}
z_k = \sum_{i=1}^M \alpha_{i,k} h_{i,k} y_{i,k} + n_k
\end{equation}
Note that the assumption of CSI at the transmitters is  important in order for the signals to add up coherently in (\ref{coherent_sum}). In principle, it can be achieved by the distributed synchronization schemes described in e.g. \cite{MudumbaiBarriacMadhow,BucklewSethares}, but may not be feasible for large sensor networks. 
However, in studies such as \cite{MudumbaiBarriacMadhow,LiDai} it has been shown in slightly different contexts that for moderate amounts of phase error much of the potential performance gains can still be achieved. 

Continuing further, we may write
$$
z_k  =  \sum_{i=1}^M \alpha_{i,k} h_{i,k} c_i x_k + \sum_{i=1}^M \alpha_{i,k} h_{i,k} v_{i,k}  + n_k 
   =  \bar{c}_k x_k + \bar{v}_k
$$
where $\bar{c}_k \equiv \sum_{i=1}^M \alpha_{i,k} h_{i,k} c_i$ and $\bar{v}_k \equiv \sum_{i=1}^M \alpha_{i,k} h_{i,k} v_{i,k}  + n_k$. Hence, we have the following linear system
\begin{equation}\begin{split}
\label{linear_system}
x_{k+1}  =  ax_k + w_k, \phantom{aa}
z_k  =  \bar{c}_k x_k + \bar{v}_k
\end{split}\end{equation}
with $\bar{v}_k$ having variance $\bar{r}_k \equiv \sum_{i=1}^M \alpha_{i,k}^2 h_{i,k}^2 \sigma_i^2 + \sigma_n^2$. Define the state estimate and  error covariance as
\begin{eqnarray*}
\hat{x}_{k+1|k} & = & \mathbb{E} \left[ x_{k+1} | \{z_0, \dots, z_k\}\right] \\
P_{k+1|k} & = &  \mathbb{E} \left[ (x_{k+1} -\hat{x}_{k+1|k})^2  | \{z_0, \dots, z_k\} \right]
\end{eqnarray*}
where again $P_{k+1|k}$ is scalar. 
Then it is well known that optimal estimation of the state $x_k$ in the minimum mean squared error (MMSE) sense can be achieved using a (in general time-varying) Kalman filter \cite{AndersonMoore}. 
Using the shorthand notation $P_{k+1}=P_{k+1|k}$, the error covariance satisfies the recursion:
\begin{equation}
\label{Riccati_eqn1}
P_{k+1}  =  a^2 P_{k} - \frac{a^2 P_{k}^2 \bar{c}_k^2}{\bar{c}_k^2 P_{k} + \bar{r}_k} + \sigma_w^2 
= \frac{a^2 P_{k} \bar{r}_k}{\bar{c}_k^2 P_{k} + \bar{r}_k} + \sigma_w^2
\end{equation}
We also remark that even if the noises are non-Gaussian, the Kalman filter is still the best \emph{linear} estimator.

\subsection{Orthogonal access scheme}
In the orthogonal access scheme each sensor transmits its measurement to the fusion center via orthogonal channels (e.g. using FDMA or CDMA), so that the fusion center receives 
$$\tilde{z}_{i,k} = \tilde{\alpha}_{i,k} \tilde{h}_{i,k} y_{i,k} + \tilde{n}_{i,k}, i=1, \dots, M$$
with the $\tilde{n}_{i,k}$'s being independent, zero mean complex Gaussian with variance $2\sigma_n^2, \forall i$. We will again assume CSI at the transmitters and use $\tilde{\alpha}_{i,k} = \alpha_{i,k} \frac{\tilde{h}_{i,k}^*}{|\tilde{h}_{i,k}|}$, with $ \alpha_{i,k} \in \mathbb{R}$. Let $h_{i,k} \equiv |\tilde{h}_{i,k}|$, $z_{i,k} \equiv \Re[\tilde{z}_{i,k}]$, $n_{i,k} \equiv \Re[\tilde{n}_{i,k}]$. The situation is then equivalent to   the linear system (using the superscript ``$o$'' to distinguish some quantities in the orthogonal scheme from the multi-access scheme):
\begin{equation*}\begin{split}
x_{k+1}  =  ax_k + w_k, \phantom{aa}
\textbf{z}_k^o  =  \bar{\textbf{C}}_k^o x_k + \bar{\textbf{v}}_k^o
\end{split}\end{equation*}
where $$\textbf{z}_k^o \equiv \left[\begin{array}{c} z_{1,k} \\ \vdots \\ z_{M,k} \end{array} \right], \bar{\textbf{C}}_k^o \equiv \left[\begin{array}{c} \alpha_{1,k} h_{1,k} c_1 \\ \vdots \\ \alpha_{M,k} h_{M,k} c_M \end{array} \right], \bar{\textbf{v}}_k^o \equiv \left[\begin{array}{c} \alpha_{1,k} h_{1,k} v_{1,k}+n_{1,k} \\ \vdots \\ \alpha_{M,k} h_{M,k} v_{M,k} + n_{M,k} \end{array} \right]$$
with the covariance of $\bar{\textbf{v}}_k^o$ being
$$\bar{\textbf{R}}_k^o \equiv \left[\begin{array}{cccc} \alpha_{1,k}^2 h_{1,k}^2 \sigma_1^2 + \sigma_n^2 & 0 & \dots & 0 \\ 0 & \alpha_{2,k}^2 h_{2,k}^2 \sigma_2^2 + \sigma_n^2  & \dots & 0 \\ \vdots & \vdots & \ddots & \vdots \\ 0 & 0 & \dots & \alpha_{M,k}^2  h_{M,k}^2 \sigma_M^2 + \sigma_n^2 \end{array} \right]$$
The state estimate and error covariance are now defined as
\begin{eqnarray*}
\hat{x}_{k+1|k}^o & = & \mathbb{E} \left[ x_{k+1} | \{\textbf{z}_0^o, \dots, \textbf{z}_k^o \}\right] \\
P_{k+1|k}^o & = &  \mathbb{E} \left[ (x_{k+1} -\hat{x}_{k+1|k}^o)^2  | \{\textbf{z}_0^o, \dots, \textbf{z}_k^o \} \right]
\end{eqnarray*}
Optimal estimation of $x_k$ in the orthogonal access scheme can also be achieved using a Kalman filter, with the error covariance now satisfying the recursion:
\begin{eqnarray*}
P_{k+1}^o & = & a^2 P_k^o - a^2 (P_k^o)^2 \bar{\textbf{C}}_k^{o^T} (\bar{\textbf{C}}_k^o P_k^o \bar{\textbf{C}}_k^{o^T} + \bar{\textbf{R}}_k^o)^{-1} \bar{\textbf{C}}_k^o + \sigma_w^2
\end{eqnarray*}
where $\bar{\textbf{C}}_k^o$ and $\bar{\textbf{R}}_k^o$ as defined above are respectively a vector and a matrix. 
To simplify the expressions, note that 
$$\bar{\textbf{C}}_k^{o^T} (\bar{\textbf{C}}_k^o P_k^o \bar{\textbf{C}}_k^{o^T} + \bar{\textbf{R}}_k^o)^{-1} \bar{\textbf{C}}_k^o = \frac{\bar{\textbf{C}}_k^{o^T} \bar{\textbf{R}}_k^{o^{-1}} \bar{\textbf{C}}_k^o}{1 + P_k^o \bar{\textbf{C}}_k^{o^T} \bar{\textbf{R}}_k^{o^{-1}} \bar{\textbf{C}}_k^o},$$
which can be shown using the matrix inversion lemma. 
Hence 
\begin{equation}
\label{Riccati_eqn2}
P_{k+1}^o
 =  \frac{a^2 P_k^o}{1 + P_k^o \bar{\textbf{C}}_k^{o^T} \bar{\textbf{R}}_k^{o^{-1}} \bar{\textbf{C}}_k^o} + \sigma_w^2
 \end{equation}
 where one can also easily compute 
$\bar{\textbf{C}}_k^{o^T} \bar{\textbf{R}}_k^{o^{-1}} \bar{\textbf{C}}_k^o = \sum_{i=1}^M \alpha_{i,k}^2 h_{i,k}^2 c_i^2/(\alpha_{i,k}^2 h_{i,k}^2 \sigma_i^2 + \sigma_n^2)$.
The advantage of the orthogonal scheme is that we do not need carrier-level synchronization among all sensors, but only require synchronization between each individual sensor and the fusion center \cite{Cui_TSP}. 

\subsection{Transmit powers}
The power $\gamma_{i,k}$ used at time $k$ by the $i$th sensor in transmitting its measurement to the fusion center is defined as
$ \gamma_{i,k} = \alpha_{i,k}^2 \mathbb{E}[y_{i,k}^2]$.
For stable scalar systems, it is well known that if $\{x_k\}$ is stationary we have 
$\mathbb{E}[x_k^2] = \frac{\sigma_w^2}{1-a^2}, \forall k$.
In both the multi-access and orthogonal schemes, the transmit powers are then:
$$\gamma_{i,k} = \alpha_{i,k}^2 \left(c_i^2 \frac{\sigma_w^2}{1-a^2} + \sigma_i^2 \right)$$

\subsection{Steady state error covariance}
In this and the next few sections we will let $\tilde{h}_{i,k} = \tilde{h}_i$ (and hence $h_{i,k} = h_i$) $, \forall k$ be time-invariant, deferring the discussion of time-varying channels until Section \ref{with_CSI_sec}. We will also assume in this case that $\alpha_{i,k} = \alpha_i, \forall k$, i.e. the amplification factors don't vary with time, and we will drop the subscript $k$ from quantities such as $\bar{c}_k$ and $\bar{r}_k$. 

From Kalman filtering theory, we know that the steady state (as $k\rightarrow \infty$) error covariance $P_\infty$ (provided it exists) in the multi-access scheme satisfies (c.f.(\ref{Riccati_eqn1}))
\begin{equation}
\label{steady_state_mac}
P_\infty =   \frac{a^2 P_\infty \bar{r}}{\bar{c}^2 P_\infty + \bar{r}}+\sigma_w^2
\end{equation}
where $\bar{r}$ and $\bar{c}$ are the time-invariant versions of $\bar{r}_k$ and $\bar{c}_k$.\footnote{The assumption of time-invariance is important. For time-varying $\bar{r}_k$ and $\bar{c}_k$, the error covariance usually will not converge to a steady state value.}
For stable systems, it is known that the steady state error covariance always exists \cite[p.77]{AndersonMoore}.  
For $\bar{c} \neq 0$,  the solution to this can be easily shown to be
\begin{equation}
\label{P_inf_scalar}
P_\infty = \frac{(a^2-1)\bar{r} + \bar{c}^2 \sigma_w^2 + \sqrt{((a^2-1)\bar{r} + \bar{c}^2 \sigma_w^2)^2 + 4 \bar{c}^2 \sigma_w^2 \bar{r} }}{2 \bar{c}^2}
\end{equation}
In the ``degenerate'' case where $\bar{c} = 0$, we have $P_\infty = \sigma_w^2/(1-a^2)$.
It will also be usful to write (\ref{P_inf_scalar}) as
\begin{equation}
\label{P_inf_alternate}
P_\infty = \frac{a^2-1 +  \sigma_w^2 S + \sqrt{(a^2-1 +  \sigma_w^2 S)^2 + 4  \sigma_w^2 S }}{2 S}
\end{equation}
with $S \equiv  \bar{c}^2 / \bar{r}$ regarded as a signal-to-noise ratio (SNR). We have the following property.
\begin{lemma}
\label{lemma1}
$P_\infty$ as defined by (\ref{P_inf_alternate}) is a decreasing function of $S$.
\end{lemma}
\begin{proof}
See the Appendix.
\end{proof}

Similarly, in the orthogonal access scheme, the steady state error covariance $P_\infty^o$ satisfies (c.f.(\ref{Riccati_eqn2}))
\begin{equation}
\label{steady_state_orth}
\begin{split}
P_\infty^o & = \frac{a^2 P_\infty^o}{1 + P_\infty^o \bar{\textbf{C}}^{o^T} \bar{\textbf{R}}^{o^{-1}} \bar{\textbf{C}}^o} + \sigma_w^2
\end{split}
\end{equation}
where $\bar{\textbf{R}}^o$ and $\bar{\textbf{C}}^o$ are the time-invariant versions of $\bar{\textbf{R}}_k^o$ and $\bar{\textbf{C}}_k^o$. 
We can easily compute $\bar{\textbf{C}}^{o^T} \bar{\textbf{R}}^{o^{-1}} \bar{\textbf{C}}^o = \sum_{i=1}^M \alpha_i^2 h_i^2 c_i^2/(\alpha_i^2 h_i^2 \sigma_i^2 + \sigma_n^2)$
with $S^o \equiv \bar{\textbf{C}}^{o^T} \bar{\textbf{R}}^{o^{-1}} \bar{\textbf{C}}^o $ regarded as a signal-to-noise ratio.
The solution to (\ref{steady_state_orth}) can then be found as 
\begin{equation}
\label{P_inf_orth}
 P_\infty^o = \frac{a^2-1+ \sigma_w^2 S^o + \sqrt{(a^2-1+ \sigma_w^2 S^o)^2 + 4\sigma_w^2 S^o}}{2S^o} 
\end{equation}
\begin{lemma}
\label{lemma2}
$P_\infty^o$ as defined by (\ref{P_inf_orth}) is a decreasing function of $S^o$
\end{lemma}
The proof is the same as that of Lemma \ref{lemma1} in the Appendix.

Comparing (\ref{P_inf_alternate}) and (\ref{P_inf_orth}) we see that the functions for $P_\infty$ and $P_\infty^o$ are of the same form, except that in
 the multi-access scheme we have 
$$S \equiv \frac{\bar{c}^2}{\bar{r}} = \frac{\left(\sum_{i=1}^M \alpha_i h_i c_i \right)^2}{\sum_{i=1}^M \alpha_i^2 h_i^2 \sigma_i^2 + \sigma_n^2}$$
and in the orthogonal scheme we have
$$S^o \equiv \bar{\textbf{C}}^{o^T} \bar{\textbf{R}}^{o^{-1}} \bar{\textbf{C}}^o = \sum_{i=1}^M \frac{\alpha_i^2 h_i^2 c_i^2}{\alpha_i^2 h_i^2 \sigma_i^2 + \sigma_n^2}$$

\subsection{Some examples of multi-access vs orthogonal access}
\label{comparison_sec}
A natural question to ask is whether one scheme always performs better than the other, e.g. whether $S \geq S^o$ given the same values for $\alpha_i, h_i, c_i, \sigma_i^2, \sigma_n^2$  are used in both expressions. We present below a number of examples to illustrate that in general this is not true. Assume for simplicity that the $\alpha_i$'s are chosen such $\alpha_i c_i$ are positive for all $i=1,\dots,M$.

1) Consider first the case when $\sigma_n^2=0$. Then we have the inequality 
$$\sum_{i=1}^M \frac{\alpha_i^2 h_i^2 c_i^2}{\alpha_i^2 h_i^2 \sigma_i^2} \geq \frac{\left(\sum_{i=1}^M \alpha_i h_i c_i \right)^2}{\sum_{i=1}^M \alpha_i^2 h_i^2 \sigma_i^2}$$ which can be shown by applying Theorem 65 of \cite{HardyLittlewoodPolya}. So when $\sigma_n^2=0$, $S^o \geq S$ and consequently $P_\infty^o$ will be smaller than $P_\infty$. The intuitive explanation for this is that if there is no noise introduced at the fusion center, then receiving the individual measurements from the sensors is better than receiving a linear combination of the measurements, see also \cite{GanHarris}. 

2) Next we consider the case when the noise variance $\sigma_n^2$ is large. We can express $S-S^o$ as
\begin{equation*}
\begin{split}
& \frac{1} {(\sum_{i=1}^M \alpha_i^2 h_i^2 \sigma_i^2 + \sigma_n^2) \prod_{i=1}^M (\alpha_i^2 h_i^2 \sigma_i^2 + \sigma_n^2)} \Big( (\sum_{i=1}^M \alpha_i h_i c_i )^2 \prod_{i=1}^M (\alpha_i^2 h_i^2 \sigma_i^2 + \sigma_n^2) \\ & - \alpha_1^2 h_1^2 c_1^2 (\sum_{i=1}^M \alpha_i^2 h_i^2 \sigma_i^2 + \sigma_n^2) \prod_{i:i\neq 1} (\alpha_i^2 h_i^2 \sigma_i^2 + \sigma_n^2)  - \dots - \alpha_M^2 h_M^2 c_M^2 (\sum_{i=1}^M \alpha_i^2 h_i^2 \sigma_i^2 + \sigma_n^2) \prod_{i:i\neq M} (\alpha_i^2 h_i^2 \sigma_i^2 + \sigma_n^2)\Big)
\end{split}
\end{equation*}
The coefficient of the $(\sigma_n^2)^M$ term in the numerator is 
$ \left(\sum_{i=1}^M \alpha_i h_i c_i\right)^2 - \alpha_1^2 h_1^2 c_1^2 - \dots - \alpha_M^2 h_M^2 c_M^2 > 0$.
For $\sigma_n^2$ sufficiently large, this term will dominate, hence $S > S^o$ and the multi-access scheme will now have smaller error covariance than the orthogonal scheme. 

3) Now we consider the ``symmetric'' situation where $\alpha_i = \alpha, c_i=c, \sigma_i^2=\sigma_v^2, h_i = h, \forall i$. Then we have 
$$S = \frac{M^2 \alpha^2 h^2 c^2}{M \alpha^2 h^2 \sigma_v^2 + \sigma_n^2} = \frac{M \alpha^2 h^2 c^2}{\alpha^2 h^2 \sigma_v^2 + \sigma_n^2/M} \textrm{ and }
S^o = \frac{M \alpha^2 h^2 c^2}{\alpha^2 h^2 \sigma_v^2 + \sigma_n^2} 
$$
Hence $S \geq S^o$, with equality only when $\sigma_n^2=0$ (or $M=1$). Thus, in the symmetric case, the multi-access scheme outperforms the orthogonal access scheme. 

4) Suppose $\sigma_n^2 \neq 0$. We wish to know whether it is always the case that $S > S^o$ for $M$ sufficiently large. The following counterexample shows that in general this assertion is false. Let $\alpha_i = 1, h_i=1, \sigma_i^2=1, \forall i$. Let $M/2$ of the sensors have $c_i = 1$, and the other $M/2$ sensors have $c_i = 2$. We find that
$$S = \frac{(M/2+M)^2}{M+\sigma_n^2} = \frac{9}{4}\frac{M}{1+\sigma_n^2/M} \textrm{ and }
S^o= \frac{M}{2}\frac{1+4}{1+\sigma_n^2} = \frac{5}{2}\frac{M}{1+\sigma_n^2}
$$
If e.g. $\sigma_n^2=1/8$, then it may be verified that $S^o>S$ for $M < 10$, $S^o=S$ for $M=10$, and $S > S^o$ for $M > 10$, so eventually the multi-access scheme outperforms the orthogonal scheme. 
On the other hand, if
$\frac{5}{2(1+\sigma_n^2)} > \frac{9}{4}$
or $\sigma_n^2 < 1/9$, we will have $S^o > S$ no matter how large $M$ is. 

\section{Asymptotic behaviour}
\label{asymptotic_sec}
Since $P_\infty$ is a decreasing function of $S$ (similar comments apply for the orthogonal scheme), increasing $S$ will provide an improvement in performance. As $S \rightarrow \infty$, we can see from (\ref{P_inf_alternate}) that $P_\infty \rightarrow \sigma_w^2$, the process noise variance. Note that unlike e.g. \cite{GastparVetterli,Cui_TSP} where the mean squared error (MSE) can be driven to zero in situations such as when there is a large number of sensors, here the lower bound $\sigma_w^2$ on performance is always strictly greater than zero. 
When the number of sensors is fixed, then it is not too difficult to show that $S$ will be bounded no matter how large (or small) one makes the $\alpha_i$'s, so getting arbitrarily close to $\sigma_w^2$ is not possible. On the other hand, if instead the number of sensors $M$ is allowed to increase, then $P_\infty \rightarrow \sigma_w^2$  as  $M \rightarrow \infty$ can be achieved in many situations, as will be shown in the following. Moreover we will  be interested in the rate at which this convergence occurs. 

In this section we will first investigate two simple strategies, 1) $\alpha_i = 1, \forall i$, and 2) $\alpha_i = 1/\sqrt{M}, \forall i$.\footnote{These strategies are similar to the case of ``equal power constraint'' and ``total power constraint'' in \cite{LiuElGamalSayeed} (also \cite{LiDai}), and various versions have also been considered in the work of \cite{GastparVetterli,Gastpar_JSAC,Xiao_coherent_TSP,Cui_TSP}, in the context of estimation of i.i.d. processes.} For the ``symmetric'' case (i.e. the parameters are the same for each sensor) we will obtain explicit asymptotic expressions. We then use these results to bound the performance in the general asymmetric case in Section \ref{general_param_sec}. Finally, we will also investigate the asymptotic performance of a simple equal power allocation scheme in Section \ref{equal_power_sec}. We note that the results in this section assume that large $M$ is possible, e.g. ability to synchronize a large number of sensors in the multi-access scheme, or the availability of a large number of orthogonal channels in the orthogonal scheme, which may not always be the case in practice. On the other hand, in numerical investigations we have found that the results derived in this section are quite accurate even for $20-30$ sensors, see Figs. \ref{asymptotic_plot1} and \ref{asymptotic_plot2} in Section \ref{numerical_sec}.

\subsection{No scaling: $\alpha_i = 1, \forall i$}
\label{no_power_control_sec}
Let $\alpha_i = 1, \forall i$, so measurements are forwarded to the fusion center without any scaling. Assume for simplicity the symmetric case, where $c_i = c, \sigma_i^2 = \sigma_v^2, h_i = h, \forall i$. 

In the multi-access scheme,  
$\bar{c} = M h c$, and $\bar{v}_k$ has variance $\bar{r} = M h^2 \sigma_v^2 + \sigma_n^2$, so that $S=\frac{M^2 h^2 c^2}{M h^2 \sigma_v^2 + \sigma_n^2}$. Since $S \rightarrow \infty$ as $M \rightarrow \infty$, we have by the previous discussion that $P_\infty \rightarrow \sigma_w^2$. 
The rate of convergence is given by the following:

\begin{lemma}
\label{asympt_lemma1}
In the symmetric multi-access scheme with $\alpha_i = 1, \forall i$,
\begin{equation}
\label{asympt_expr_1}
P_\infty = \sigma_w^2 + \frac{a^2 \sigma_v^2}{c^2}\frac{1}{M} + O\left(\frac{1}{M^2}\right)
\end{equation}
as $M \rightarrow \infty$. 
\end{lemma} 
\begin{proof}
See the Appendix.
\end{proof}

Thus the steady state error covariance for the multi-access scheme converges to the process noise variance $\sigma_w^2$, at a rate of $1/M$. This result matches the rate of $1/M$ achieved for estimation of i.i.d. processes using multi-access schemes, e.g. \cite{GastparVetterli,LiuElGamalSayeed}.

In the orthogonal scheme we have
$ S^o = \frac{M h^2 c^2}{h^2 \sigma_v^2 + \sigma_n^2}$, so $S^o \rightarrow \infty$ as $M \rightarrow \infty$ also. By similar calculations to the proof of Lemma \ref{asympt_lemma1} we find that as $M \rightarrow \infty$
\begin{equation}
\label{asympt_expr_2b}
 P_\infty^o = \sigma_w^2 + \frac{a^2 (h^2 \sigma_v^2+\sigma_n^2)}{h^2 c^2}\frac{1}{M} + O\left(\frac{1}{M^2}\right) = \sigma_w^2 + \frac{a^2 (\sigma_v^2+\sigma_n^2/h^2)}{c^2}\frac{1}{M} + O\left(\frac{1}{M^2}\right).
\end{equation}
Therefore, the steady state error covariance again converges to $\sigma_w^2$ at a rate of $1/M$, but the constant $\frac{a^2 (\sigma_v^2+\sigma_n^2/h^2)}{c^2}$ in front is larger. This agrees with example 3) of Section \ref{comparison_sec} that in the symmetric situation the multi-access scheme will perform better than the orthogonal scheme.

\subsection{Scaling $\alpha_i = 1/\sqrt{M}, \forall i$}
\label{1/M_sec}
In the previous case with $\alpha_i = 1, \forall i$, the power received at the fusion center will grow unbounded as $M \rightarrow \infty$. Suppose instead we let $\alpha_i = 1/\sqrt{M}, \forall i$, which will keep the power received at the fusion center bounded (and is constant in the symmetric case), while the transmit power used by each sensor will tend to zero as $M \rightarrow \infty$. Again assume for simplicity that $c_i = c, \sigma_i^2 = \sigma_v^2, h_i = h, \forall i$. 

In the multi-access scheme we now have $S = \frac{M h^2 c^2}{h^2 \sigma_v^2 + \sigma_n^2},$ 
so that as $M \rightarrow \infty$,
\begin{equation}
\label{asympt_expr_2} P_\infty =  \sigma_w^2 + \frac{a^2 (\sigma_v^2 + \sigma_n^2/h^2)}{c^2}\frac{1}{M} + O\left(\frac{1}{M^2}\right).
\end{equation}
Thus we again have the steady state error covariance converging to the process noise variance $\sigma_w^2$  at a rate of $1/M$. In fact, we see that this is the same expression as (\ref{asympt_expr_2b}) in the orthogonal scheme, but where we were using $\alpha_i = 1, \forall i$. The difference here is that this performance can be achieved even when the transmit power used by each individual sensor will \emph{decrease to zero} as the number of sensors increases, which could be quite desirable in power constrained environments such as wireless sensor networks. 
For i.i.d. processes, this somewhat surprising behaviour when the total received power is bounded has also been observed \cite{Gastpar_JSAC,LiuElGamalSayeed}. 

In the orthogonal scheme we have $S^o = \frac{h^2 c^2}{h^2 \sigma_v^2/M + \sigma_n^2}$, and we note that now $S^o$ is bounded even as $M \rightarrow \infty$, so $P_\infty^o$ cannot  converge to $\sigma_w^2$ as $M \rightarrow \infty$. 
For a more precise expression, we can show by similar computations to the proof of Lemma \ref{asympt_lemma1} that for large $M$,
\begin{equation}
\label{P_inf_orthogonal}
\begin{split}
P_\infty^o  = & \frac{(a^2-1)\sigma_n^2 + h^2 c^2 \sigma_w^2 + \sqrt{(a^2-1)^2 \sigma_n^4 + 2(a^2+1)\sigma_n^2 h^2 c^2 \sigma_w^2 + h^4 c^4 \sigma_w^4}}{2 h^2 c^2} \\
&  + \left[ \frac{(a^2-1)\sigma_v^2}{2 c^2}+ \frac{(a^2+1)h^4 \sigma_v^2 c^2 \sigma_w^2+(a^2-1)^2 \sigma_n^2 h^2 \sigma_v^2}{2 h^2 c^2 \sqrt{(a^2-1)^2 \sigma_n^4 + 2(a^2+1)\sigma_n^2 h^2 c^2 \sigma_w^2 + h^4 c^4 \sigma_w^4}} \right] \frac{1}{M} + O\left(\frac{1}{M^2}\right)
\end{split}
\end{equation}
Noting that 
$\frac{(a^2-1)\sigma_n^2 + h^2 c^2 \sigma_w^2 + \sqrt{(a^2-1)^2 \sigma_n^4 + 2(a^2+1)\sigma_n^2 h^2 c^2 \sigma_w^2 + h^4 c^4 \sigma_w^4}}{2 h^2 c^2} > \sigma_w^2$, the steady state error covariance will converge as $M \rightarrow \infty$ to a value strictly greater than $\sigma_w^2$, though the convergence is still at a rate $1/M$. 
 Analogously, for i.i.d. processes it has been shown that in the orthogonal scheme the MSE does not go to zero as $M \rightarrow \infty$  when the total power used is bounded \cite{Cui_TSP}. 

\subsection{General parameters}
\label{general_param_sec}
The behaviour shown in the two previous cases can still hold under more general conditions on $c_i$, $\sigma_i^2$ and $h_i$. Suppose for instance that they can be bounded from both above and below, i.e.
$0 < c_{min} \leq |c_i| \leq c_{max} < \infty$, $0 < \sigma_{min}^2 \leq \sigma_i^2 \leq \sigma_{max}^2 < \infty$, $0 < h_{min} \leq h_i \leq h_{max} < \infty, \forall i$.
We have the following:
\begin{lemma}
\label{asympt_lemma2}
In the general multi-access scheme, as $M \rightarrow \infty$, using either no scaling of measurements, or scaling of measurements by $1/\sqrt{M}$, results in
$$P_\infty = \sigma_w^2 + O \left(\frac{1}{M} \right)$$
In the general orthogonal scheme, using no scaling of measurements results in
$$P_\infty^o = \sigma_w^2 + O \left(\frac{1}{M} \right)$$
as $M \rightarrow \infty$, but $P_\infty^o$ does not converge to a limit (in general) as $M \rightarrow \infty$ when measurements are scaled by $1/\sqrt{M}$.
\end{lemma}

\begin{proof}
See the Appendix.
\end{proof}

\subsection{Asymptotic behaviour under equal power allocation}
\label{equal_power_sec}
When the parameters are asymmetric, the above rules will in general allocate different powers to the individual sensors. Another simple alternative is to use equal power allocation. Recall that the transmit power used by each sensor is 
$\gamma_{i} = \alpha_{i}^2 \left(c_i^2 \frac{\sigma_w^2}{1-a^2} + \sigma_i^2 \right)$.
If we allocate power $\gamma$ to each sensor, i.e. $\gamma_i = \gamma, \forall i$, then 
\begin{equation}
\label{equal_power_alloc_1}
\alpha_i = \sqrt{\frac{\gamma(1-a^2)}{c_i^2 \sigma_w^2 + \sigma_i^2 (1-a^2)}}
\end{equation}
If instead the total power $\gamma_{total}$ is to be shared equally amongst sensors, then $\gamma_i = \gamma_{total}/M, \forall i$, and 
\begin{equation}
\label{equal_power_alloc_2}
\alpha_i = \sqrt{\frac{\gamma_{total}(1-a^2)}{M\left(c_i^2 \sigma_w^2 + \sigma_i^2(1-a^2)\right)}}
\end{equation}
Asymptotic results under equal power allocation are quite similar to Section \ref{general_param_sec}, namely: 
\begin{lemma}
\label{asympt_lemma3}
In the general multi-access scheme, as $M \rightarrow \infty$, using the equal power allocation (\ref{equal_power_alloc_1}) or (\ref{equal_power_alloc_2}) results in
$$P_\infty = \sigma_w^2 + O \left(\frac{1}{M} \right)$$
In the general orthogonal scheme, using the equal power allocation (\ref{equal_power_alloc_1}) results in
$$P_\infty^o = \sigma_w^2 + O \left(\frac{1}{M} \right)$$
as $M \rightarrow \infty$, but $P_\infty^o$ does not converge to a limit as $M \rightarrow \infty$ when using the power allocation (\ref{equal_power_alloc_2}).
\end{lemma}

\begin{proof}
See the Appendix
\end{proof}

\subsection{Remarks}
1) Most of the previous policies in this section give a convergence rate of $1/M$. We might wonder whether one can achieve an even better rate (e.g. $1/M^2$) using other choices for $\alpha_i$, though the answer turns out to be no. To see this, following \cite{GastparVetterli}, consider the ``ideal'' case where sensor measurements are received perfectly at the fusion center, and which mathematically corresponds to the orthogonal scheme with $\sigma_n^2=0, \alpha_i=1, h_i=1, \forall i$. This idealized situation provides a lower bound on the achievable error covariance. We will have
$S^o = \sum_{i=1}^M c_i^2/\sigma_i^2$,
which can then be used to show that $P_\infty^o$ converges to $\sigma_w^2$ at the rate $1/M$. Hence $1/M$ is the best rate that can be achieved with any coded/uncoded scheme. 

2) In the previous derivations we have not actually used the assumption that $|a| < 1$, so the results in Sections \ref{no_power_control_sec} - \ref{general_param_sec} will hold even when the system is unstable (assuming $\bar{C} \neq 0$). However for unstable systems, $\mathbb{E}[x_k^2]$ becomes unbounded as $k \rightarrow \infty$, so if the $\alpha_{i,k}$'s are time invariant, then more and more power is used by the sensors as time passes. If the application is a wireless sensor network where power is limited, then the question is whether one can choose these $\alpha_{i,k}$'s such that \emph{both} the power used by the sensors  and the error covariances will be bounded for all times. Now if there is no noise at the fusion center, i.e. $n_k=0$, then a simple scaling of the measurements at the individual sensors will work. But when $n_k \neq 0$, as will usually be the case in analog forwarding, we have not been able to find a scheme which can achieve this. Note however that 
for unstable systems, asymptotic results are of mathematical interest only. In practice, in most cases, we will 
be interested in finite horizon results for unstable systems where the system states and measurements can take 
on large values but are still bounded. In such  finite horizon situations, one can perform optimum power allocation at each time step similar to Section \ref{with_CSI_sec} but for a finite number of time steps, or use a finite horizon dynamic programming approach similar to Section \ref{DP_sec}. However these problems will not be addressed in the current paper. 

\section{Optimal power allocation}
\label{optim_sec}
When there are a large number of sensors, one can use simple strategies such as $\alpha_i = 1/\sqrt{M}, \forall i$, or the equal power allocation (\ref{equal_power_alloc_2}), which will both give a convergence of the steady state error covariance to $\sigma_w^2$ at a rate of $1/M$ in the multi-access scheme, while bounding the total power used by all the sensors. However when the number of sensors is small, one may perhaps do better with different choices of the $\alpha_i$'s. 
In this section we will study some relevant power allocation problems. These are considered first for static channels in the multi-access and orthogonal schemes, in Sections \ref{mac_optim_sec} and \ref{orth_optim_sec} respectively. Some features of the solutions to these optimization problems are discussed in Section \ref{optim_interpretation}. These results are then extended to fading channels with channel state information (CSI) and fading channels without CSI in Sections \ref{with_CSI_sec} and \ref{no_CSI_sec} respectively.

\subsection{Optimization problems for multi-access scheme}
\label{mac_optim_sec}
\subsubsection{Minimizing sum power}
\label{min_power_sec}
One possible formulation is to minimize the sum of transmit powers used by the sensors subject to a bound $D$ on the steady state error covariance. More formally, the problem is
\begin{equation*}
\begin{split}
& \min \sum_{i=1}^M \gamma_i  = \sum_{i=1}^M \alpha_i^2 \left( \frac{c_i^2 \sigma_w^2}{1-a^2} + \sigma_i^2 \right)\\
& \mbox{ subject to } P_\infty \leq D
\end{split}
\end{equation*}
with $P_\infty$ given by (\ref{P_inf_scalar}).
Some straightforward manipulations show that the constraint can be simplified to
\begin{equation}
\label{constraint_eqn}
 \bar{r} \left( a^2 D + \sigma_w^2 - D \right) + \bar{c}^2 D (\sigma_w^2-D) \leq 0
\end{equation}
i.e.
$$ \left(\sum_{i=1}^M \alpha_i^2 h_i^2 \sigma_i^2 + \sigma_n^2\right)  \left( a^2 D + \sigma_w^2 - D \right)  + \left(\sum_{i=1}^M \alpha_i h_i c_i\right)^2 D (\sigma_w^2-D) \leq 0$$ 
Now define
$s = h_1 c_1 \alpha_1 + \dots + h_M c_M \alpha_M$. Then the optimization problem becomes
\begin{equation}
\label{P1s}
\begin{split}
 \min_{\alpha_1,\dots,\alpha_M,s} & \sum_{i=1}^M \alpha_i^2 \left( \frac{c_i^2 \sigma_w^2}{1-a^2} + \sigma_i^2 \right)\\
 \mbox{ subject to } & \left( \sum_{i=1}^M \alpha_i^2 h_i^2 \sigma_i^2 + \sigma_n^2 \right) \left(a^2 D + \sigma_w^2 - D\right) \leq s^2 D(D-\sigma_w^2) \textrm{ and } s = \sum_{i=1}^M h_i c_i \alpha_i 
\end{split}
\end{equation}
Before continuing further, let us first determine some upper and lower bounds on $D$. From  Section \ref{asymptotic_sec}, a lower bound is $D \geq \sigma_w^2$, the process noise variance. For an upper bound, suppose $\bar{c}=0$ so we don't have any information about $x_k$. Since we are assuming the system is stable, one can still achieve an error covariance of $\frac{\sigma_w^2}{1-a^2}$ (just let $\hat{x}_k=0,\forall k$), so $D \leq \frac{\sigma_w^2}{1-a^2}$. Hence in problem (\ref{P1s}) both $D-\sigma_w^2$ and $a^2 D + \sigma_w^2 - D$ are positive quantities. 

To reduce the amount of repetition in later sections, consider the slightly more general problem
\begin{equation}
\label{P1}
\begin{split}
 \min_{\alpha_1,\dots,\alpha_M,s} & \sum_{i=1}^M \alpha_i^2 \kappa_i\\
 \mbox{ subject to } & \left( \sum_{i=1}^M \alpha_i^2 \tau_i + \sigma_n^2 \right)x  \leq s^2 y \textrm{ and }  s = \sum_{i=1}^M  \alpha_i \rho_i
\end{split}
\end{equation}
where $x>0, y>0, \kappa_i > 0, \rho_i\in \mathbb{R}, \tau_i>0,  i = 1, \dots, M$ are constants. 
In the context of (\ref{P1s}), $x=a^2 D + \sigma_w^2 - D$, $y=D(D-\sigma_w^2)$, $\rho_i=h_i c_i, \; \tau_i=h_i^2 \sigma_i^2$ and 
$\kappa_i=\left( \frac{c_i^2 \sigma_w^2}{1-a^2} + \sigma_i^2 \right)$ for $i=1,\dots,M$.

The objective function of problem (\ref{P1}) is clearly convex. Noting that $\tau_i, \sigma_n^2, x$ and $y$ are all positive, the set of points satisfying $\left( \sum_{i=1}^M \tau_i \alpha_i^2  + \sigma_n^2 \right)x  = y s^2 $ is then a quadric surface that consists of two pieces, corresponding to $s>0$ and $s<0$.\footnote{In three dimensions this surface corresponds to a ``hyperboloid of two sheets''.} Furthermore, the set of points satisfying $\left( \sum_{i=1}^M \tau_i \alpha_i^2  + \sigma_n^2 \right)x  \leq y s^2 $ and $s>0$, and the set of points satisfying $\left( \sum_{i=1}^M \tau_i \alpha_i^2  + \sigma_n^2 \right)x  \leq y s^2 $ and $s<0$, are both known to be convex sets, see e.g. Prop. 15.4.7 of \cite{Berger2}.  Hence the parts of the feasible region corresponding to $s>0$ and $s < 0$ are both convex, and the global solution can be efficiently obtained numerically.  Furthermore, following similar steps to \cite{Xiao_coherent_TSP}, a solution in (mostly) closed form can actually be obtained. We omit the derivations but shall  summarise what is required.

One first solves numerically for $\lambda$ the equation
\begin{equation*}
\sum_{i=1}^M \frac{\lambda \rho_i^2}{\kappa_i + \lambda \tau_i x} = \frac{1}{y}
\end{equation*}
 Since the left hand side is increasing with $\lambda$ solutions to this equation will be unique provided it exists. Taking limits as $\lambda \rightarrow \infty$, we see that a solution exists if and only if 
\begin{equation}
\label{feasibility_condition}
\sum_{i=1}^M \frac{\rho_i^2}{\tau_i} > \frac{x}{y}
\end{equation}
Equation (\ref{feasibility_condition}) thus provides a feasibility check for the optimization problem (\ref{P1}). In the context of 
(\ref{P1s}), one can easily derive that 
(\ref{feasibility_condition}) implies $\sum_{i=1}^M \frac{c_i^2}{\sigma_i^2} > \frac{a^2 D +\sigma_w^2-D}{D(D-\sigma_w^2)}$, which indicates 
that the sum of the sensor signal to noise ratios must be greater than a threshold (dependent on the error covariance 
threshold $D$) for the optimization problem 
(\ref{P1s}) to be feasible.

Next, we compute $\mu$ from 
\begin{equation*}
\begin{split}
\mu^2 & =  \sigma_n^2 x \left( \sum_{i=1}^M  \frac{\rho_i^2 \kappa_i}  {4 \lambda(\kappa_i + \lambda \tau_i x)^2 } \right)^{-1}
\end{split}
\end{equation*}

Finally we obtain the optimal $\alpha_i$'s (denoted by $\alpha_i^*$)
\begin{equation}
\label{alpha_P1}
 \alpha_i^* = \frac{\mu \rho_i}{2(\kappa_i + \lambda \tau_i x)}, i = 1, \dots, M.
\end{equation}
with the resulting powers
$$\gamma_{i} = \alpha_i^{*2} \kappa_i = \alpha_i^{*2} \left(c_i^2 \frac{\sigma_w^2}{1-a^2} + \sigma_i^2 \right), i = 1, \dots, M$$
Note that depending on whether we choose $\mu$ to be positive or negative, two different sets of $\alpha_i^*$'s will be obtained, one of which is the negative of the other, though the $\gamma_i$'s and hence the optimal value of the objective function remains the same. 

Another interesting relation that can be shown (see \cite{Xiao_coherent_TSP}) is that the optimal sum power satisfies
\begin{equation}
\label{mac_sum_power_relation}
\gamma_{total}^* = \sum_{i=1}^M \alpha_i^{*2} \kappa_i = \lambda \sigma_n^2 x
\end{equation}
This relation will be useful in obtaining an analytic solution to problem (\ref{P2}) next. 

\subsubsection{Minimizing error covariance}
\label{min_distortion_sec}
A related problem is to minimize the steady state error covariance subject to a sum power constraint $\gamma_{total}$. Formally, this is 
\begin{equation*}
\begin{split}
& \min P_\infty  \\
& \mbox{ subject to }  \sum_{i=1}^M \alpha_i^2 \left( \frac{c_i^2 \sigma_w^2}{1-a^2} + \sigma_i^2 \right) \leq \gamma_{total}
\end{split}
\end{equation*}
with $P_\infty$ again given by (\ref{P_inf_scalar}). For this problem, the feasible region is clearly convex, but the objective function is complicated. To simplify the objective, recall from Lemma \ref{lemma1} that $P_\infty$ is a decreasing function of $S=\bar{c}^2/\bar{r}$. Thus maximizing $ \bar{c}^2/\bar{r}$ (or minimizing $ \bar{r}/\bar{c}^2$) is equivalent to minimizing $P_\infty$, which  has the interpretation that maximizing the SNR minimizes $P_\infty$. Hence the problem is equivalent to 
\begin{equation*}
\begin{split}
\min_{\alpha_1,\dots,\alpha_M,s} & \frac{\sum_{i=1}^M \alpha_i^2 h_i^2 \sigma_i^2 + \sigma_n^2}{s^2} \\
 \mbox{ subject to } &   \sum_{i=1}^M \alpha_i^2 \left( \frac{c_i^2 \sigma_w^2}{1-a^2} + \sigma_i^2 \right)  \leq \gamma_{total}  \textrm{ and } s = \sum_{i=1}^M h_i c_i \alpha_i 
\end{split}
\end{equation*}
We again introduce a more general problem
\begin{equation}
\label{P2}
\begin{split}
\min_{\alpha_1,\dots,\alpha_M,s} & \frac{\sum_{i=1}^M \alpha_i^2 \tau_i + \sigma_n^2}{s^2} \\
 \mbox{ subject to } &   \sum_{i=1}^M \alpha_i^2 \kappa_i  \leq \gamma_{total}  \textrm{ and }
  s = \sum_{i=1}^M \alpha_i \rho_i
\end{split}
\end{equation}
with $x>0, y>0, \kappa_i > 0, \rho_i\in\mathbb{R}, \tau_i>0,  i = 1, \dots, M$ being constants. 
The objective function is still non-convex, however by making use of the properties of the analytical solution to problem (\ref{P1}), such as the relation (\ref{mac_sum_power_relation}), an analytical solution to problem (\ref{P2}) can also be obtained.  The optimal $\alpha_i$'s can be shown to satisfy:
\begin{equation}
\label{alpha_P2}
\alpha_i^{*2} = \gamma_{total} \left(\sum_{j=1}^M \frac{\rho_j^2}{(\kappa_j + \gamma_{total} \frac{\tau_j}{\sigma_n^2})^2}\kappa_j \right)^{-1} \frac{\rho_i^2}{(\kappa_i+\gamma_{total} \frac{\tau_i}{\sigma_n^2})^2}\kappa_i
\end{equation}
The details on obtaining this solution are similar to \cite{Xiao_coherent_TSP} and omitted.

\subsection{Optimization problems for orthogonal access scheme}
\label{orth_optim_sec}
\subsubsection{Minimizing sum power}
The corresponding problem of minimizing the sum power in the orthogonal scheme is
\begin{equation*}
\begin{split}
& \min \sum_{i=1}^M \gamma_i  = \sum_{i=1}^M \alpha_i^2 \left( \frac{c_i^2 \sigma_w^2}{1-a^2} + \sigma_i^2 \right)\\
& \mbox{ subject to } P_\infty^o \leq D
\end{split}
\end{equation*}
with $P_\infty^o$ now given by (\ref{P_inf_orth}). 
By a rearrangement of the constraint, this can be shown to be equivalent to 
\begin{equation}
\label{P3s}
\begin{split}
& \min_{\alpha_1^2, \dots, \alpha_M^2}  \sum_{i=1}^M \alpha_i^2 \left( \frac{c_i^2 \sigma_w^2}{1-a^2} + \sigma_i^2 \right)\\
& \mbox{ subject to } \sum_{i=1}^M \frac{\alpha_i^2 h_i^2 c_i^2}{\alpha_i^2 h_i^2 \sigma_i^2 + \sigma_n^2} \geq \frac{a^2 D+ \sigma_w^2-D}{D(D-\sigma_w^2)}
\end{split}
\end{equation}
Note that in contrast to the multi-access scheme, we now write the minimization over $\alpha_i^2$ rather than $\alpha_i$. 
Since each of the functions 
$$ \frac{-\alpha_i^2 h_i^2 c_i^2}{\alpha_i^2 h_i^2 \sigma_i^2 + \sigma_n^2} = \frac{-c_i^2}{\sigma_i^2}+\frac{\sigma_n^2 c_i^2/\sigma_i^2}{\alpha_i^2 h_i^2 \sigma_i^2 + \sigma_n^2}$$
is convex in $\alpha_i^2$, the problem will be a convex optimization problem in $(\alpha_1^2, \dots, \alpha_M^2)$. Note that without further restrictions on $\alpha_i$ we will get $2^M$ solutions with the same values of the objective function, corresponding to the different choices of positive and negative signs on the $\alpha_i$'s. This is in contrast to the multi-access scheme where there were two sets of solutions. For simplicity we can take the solution corresponding to all $\alpha_i \geq 0$.\footnote{In general this is not possible in the multi-access scheme. For instance, if we have two sensors with $c_1$ being positive and $c_2$ negative, the optimal solution will involve $\alpha_1$ being positive and $\alpha_2$ negative, or vice versa. Restricting both $\alpha_i$'s to be positive in the multi-access scheme will result in a sub-optimal solution.} 

An analytical solution can also be obtained. To reduce repetition in later sections, consider the more general problem 
\begin{equation}
\label{P3}
\begin{split}
& \min_{\alpha_1^2, \dots, \alpha_M^2}  \sum_{i=1}^M \alpha_i^2 \kappa_i \\
& \mbox{ subject to } \sum_{i=1}^M \frac{\alpha_i^2 \rho_i^2}{\alpha_i^2 \tau_i + \sigma_n^2} \geq \frac{x}{y}
\end{split}
\end{equation}
where $x>0, y>0, \kappa_i > 0, \rho_i\in \mathbb{R}, \tau_i>0,  i = 1, \dots, M$ are constants and
have similar interpretations as in Section \ref{min_power_sec}. 
Since the derivation of the analytical solution is similar to that found in \cite{Cui_TSP} (though what they regard as $\alpha_k$ is $\alpha_i^2$ here), it will be omitted and we will only present the solution. 

Firstly, the problem will be feasible if and only if 
$$ \sum_{i=1}^M \frac{\rho_i^2}{\tau_i} > \frac{x}{y}$$
Interestingly, this is the same  as the feasibility condition (\ref{feasibility_condition}) for problem (\ref{P1}) in the multi-access scheme, indicating that the total SNR for the sensor measurements must be greater than a certain threshold (dependent on $D$). 
The optimal $\alpha_i$'s satisfy
\begin{equation}
\label{alpha_P3}
\alpha_i^{*2} =   \frac{1}{\tau_i} \left(\sqrt{\frac{\lambda \rho_i^2 \sigma_n^2}{\kappa_i}}-\sigma_n^2\right)^+  
\end{equation}
where $(x)^+$ is the function that is equal to $x$ when $x$ is positive, and zero otherwise. 
To determine $\lambda$, now assume that the sensors are ordered such that
$$\frac{\rho_1^2}{\kappa_1} \geq \dots \geq \frac{\rho_M^2}{\kappa_M}.$$
Note that in the context of problem (\ref{P3s}), $\frac{\rho_i^2}{\kappa_i}= \frac{h_i^2}{\sigma_w^2/(1-a^2) + \sigma_i^2/c_i^2}$. 
Clearly, this ordering favours the sensors with better channels and higher  measurement quality.  
Then the optimal values of $\alpha_i^2$ (and hence $\alpha_i^*$) can also be expressed as
$$\alpha_i^{*2} = \left\{ \begin{array}{ccl}
  \frac{1}{\tau_i} (\sqrt{\frac{\lambda \rho_i^2 \sigma_n^2}{\kappa_i}}-\sigma_n^2) & , & i \leq M_1 \\
  0 & , & \textrm{otherwise}
  \end{array} \right.
$$
where 
$$\sqrt{\lambda} = \frac{\sum_{i=1}^{M_1}\frac{|\rho_i|}{\tau_i}\sqrt{\kappa_i \sigma_n^2}}{\sum_{i=1}^{M_1}\frac{\rho_i^2}{\tau_i}-\frac{x}{y}}$$ 
and the number of sensors which are active, $M_1$ (which can be shown to be unique \cite{Xiao_TSP}), satisfies 
$$ \sum_{i=1}^{M_1} \frac{\rho_i^2}{\tau_i} - \frac{x}{y} \geq 0, \frac{\sum_{i=1}^{M_1}\frac{|\rho_i|}{\tau_i}\sqrt{\kappa_i \sigma_n^2}}{\sum_{i=1}^{M_1}\frac{\rho_i^2}{\tau_i}-\frac{x}{y}} \sqrt{\frac{\rho_{M_1}^2 \sigma_n^2}{\kappa_{M_1}}} - \sigma_n^2 > 0 \textrm{ and } \frac{\sum_{i=1}^{M_1+1}\frac{|\rho_i|}{\tau_i}\sqrt{\kappa_i \sigma_n^2}}{\sum_{i=1}^{M_1+1}\frac{\rho_i^2}{\tau_i}-\frac{x}{y}} \sqrt{\frac{\rho_{M_1+1}^2 \sigma_n^2}{\kappa_{M_1+1}}} - \sigma_n^2 \leq 0$$

\subsubsection{Minimizing error covariance}
The corresponding problem of minimizing the error covariance in the orthogonal scheme is equivalent to 
\begin{equation*}
\begin{split}
& \min_{\alpha_1^2, \dots, \alpha_M^2} - \sum_{i=1}^M  \frac{\alpha_i^2 h_i^2 c_i^2}{\alpha_i^2 h_i^2 \sigma_i^2 + \sigma_n^2} \\
& \mbox{ subject to }\sum_{i=1}^M \alpha_i^2 \left( \frac{c_i^2 \sigma_w^2}{1-a^2} + \sigma_i^2 \right) \leq \gamma_{total} 
\end{split}
\end{equation*}
which is again a convex problem in $(\alpha_1^2, \dots, \alpha_M^2)$. For an analytical solution \cite{Cui_TSP}, consider a more general problem
\begin{equation}
\label{P4}
\begin{split}
& \min_{\alpha_1^2, \dots, \alpha_M^2} - \sum_{i=1}^M \frac{\alpha_i^2 \rho_i^2}{\alpha_i^2 \tau_i + \sigma_n^2} \\
& \mbox{ subject to }   \sum_{i=1}^M \alpha_i^2 \kappa_i  \leq \gamma_{total}
\end{split}
\end{equation}
where $x>0, y>0, \kappa_i > 0, \rho_i\in\mathbb{R}, \tau_i>0,  i = 1, \dots, M$ are constants.
Then the optimal $\alpha_i$'s satisfy
\begin{equation}
\label{alpha_P4}
 \alpha_i^{*2} = \frac{1}{\tau_i} \left(\sqrt{\frac{ \rho_i^2 \sigma_n^2}{\lambda \kappa_i}}-\sigma_n^2\right)^+ .
\end{equation}
Assuming that the sensors are ordered so that
$$\frac{\rho_1^2}{\kappa_1} \geq \dots \geq \frac{\rho_M^2}{\kappa_M}$$
the optimal values of $\alpha_i^2$ to problem (\ref{P4}) can also be expressed as
$$\alpha_i^{*2} = \left\{ \begin{array}{ccl}
  \frac{1}{\tau_i} (\sqrt{\frac{ \rho_i^2 \sigma_n^2}{\lambda \kappa_i}}-\sigma_n^2) & , & i \leq M_1 \\
  0 & , & \textrm{otherwise}
  \end{array} \right.
$$
where 
$$\frac{1}{\sqrt{\lambda}} = \frac{\gamma_{total} + \sum_{i=1}^{M_1}\frac{\kappa_i}{\tau_i}\sigma_n^2}{\sum_{i=1}^{M_1}\frac{|\rho_i|}{\tau_i}\sqrt{\kappa_i \sigma_n^2}}$$ 
and the number of sensors which are active, $M_1$ (which is again unique), satisfies 
$$\frac{\gamma_{total} + \sum_{i=1}^{M_1}\frac{\kappa_i}{\tau_i}\sigma_n^2}{\sum_{i=1}^{M_1}\frac{|\rho_i|}{\tau_i}\sqrt{\kappa_i \sigma_n^2}} \sqrt{\frac{\rho_{M_1}^2 \sigma_n^2}{\kappa_{M_1}}} - \sigma_n^2 > 0 \textrm{ and } \frac{\gamma_{total} + \sum_{i=1}^{M_1+1}\frac{\kappa_i}{\tau_i}\sigma_n^2}{\sum_{i=1}^{M_1+1}\frac{|\rho_i|}{\tau_i}\sqrt{\kappa_i \sigma_n^2}} \sqrt{\frac{\rho_{M_1+1}^2 \sigma_n^2}{\kappa_{M_1+1}}} - \sigma_n^2 \leq 0$$

\subsection{Remarks}
\label{optim_interpretation}
1)
In the orthogonal scheme, the solutions of the optimization problems (\ref{P3}) and (\ref{P4}) take the form (\ref{alpha_P3}) and (\ref{alpha_P4}) respectively. These expressions are reminiscent of the ``water-filling'' solutions in wireless communications, where only sensors of sufficiently high quality measurements will be allocated power, while sensors with lower quality measurements are turned off. On the other hand, the solutions for problems (\ref{P1}) and (\ref{P2}) have the form (\ref{alpha_P1}) and (\ref{alpha_P2}) respectively, which indicates that all sensors will get allocated some non-zero power when we perform the optimization. The intuition behind this is that in the multi-access scheme some ``averaging'' can be done when measurements are added together, which can reduce the effects of noise and improve performance, while this can't be done in the orthogonal scheme so that turning off low quality sensors will save power.

2)
The four optimization problems we consider (problems (\ref{P1}), (\ref{P2}), (\ref{P3}) and (\ref{P4})) have analytical solutions, and can admit distributed implementations, which may be important in large sensor networks. For problem (\ref{P1}) the fusion center can calculate the values $\lambda$ and $\mu$ and broadcast them to all sensors, and for problem (\ref{P2}) the fusion center can calculate and broadcast the quantity
$\left(\sum_{j=1}^M \frac{\rho_j^2}{(\kappa_j + \gamma_{total} \tau_j/\sigma_n^2)^2}\kappa_j \right)^{-1}$
to all sensors. The sensors can then use these quantities and their local information to compute the optimal $\alpha_i$'s, see \cite{Xiao_coherent_TSP}. 
For problems (\ref{P3}) and (\ref{P4}), the fusion center can compute and broadcast the quantity $\lambda$ to all sensors, which can then determine their optimal $\alpha_i$'s using $\lambda$ and their local information, see \cite{Cui_TSP}.

\subsection{Fading channels with CSI}
\label{with_CSI_sec}
We will now consider channel gains that are randomly time-varying. 
In this section we let both the sensors and fusion center have channel state information (CSI), so that the $h_{i,k}$'s are known, while Section \ref{no_CSI_sec} considers fading channels without CSI. We now also allow the amplification factors $\alpha_{i,k}$ to be time-varying.  

\subsubsection{Multi-access}
Recall from (\ref{Riccati_eqn1}) that the Kalman filter recursion for the error covariances is 
$$P_{k+1}  =  \frac{a^2 P_{k} \bar{r}_k}{\bar{c}_k^2 P_{k} + \bar{r}_k} + \sigma_w^2
$$
where $\bar{c}_k \equiv \sum_{i=1}^M \alpha_{i,k} h_{i,k} c_i$ and $\bar{r}_k \equiv \sum_{i=1}^M \alpha_{i,k}^2 h_{i,k}^2 \sigma_i^2 + \sigma_n^2$. 

One way in which we can formulate an optimization problem is to minimize the sum of powers used at each time instant, subject to $P_{k+1|k}\leq D$ at all time instances $k$. That is, for all $k$, we want to solve
\begin{equation}
\label{optim_prob_fading}
\begin{split}
& \min \sum_{i=1}^M \gamma_{i,k}  = \sum_{i=1}^M \alpha_{i,k}^2 \left( \frac{c_i^2 \sigma_w^2}{1-a^2} + \sigma_i^2 \right)\\
& \mbox{ subject to } P_{k+1} = \frac{a^2 P_{k} \bar{r}_k}{\bar{c}_k^2 P_{k} + \bar{r}_k} + \sigma_w^2 \leq D
\end{split}
\end{equation}
The constraint can be rearranged to be equivalent to
$$ \bar{r}_k \left( a^2 P_k + \sigma_w^2 - D \right) + \bar{c}_k^2 P_k (\sigma_w^2-D) \leq 0$$ 
which looks rather similar to (\ref{constraint_eqn}). In fact, once we've solved the problem (\ref{optim_prob_fading}) at an initial time instance, e.g. $k=1$, then $P_2 = D$ is satisfied, so that further problems become essentially identical to what was solved in Section \ref{min_power_sec}. Therefore, the only slight difference is in the initial optimization problem, though this is also covered by the general problem (\ref{P1}). 

Another possible optimization problem is to minimize $P_{k+1|k}$ at each time instant subject to a sum power constraint $\gamma_{total}$ at each time $k$, i.e.
\begin{equation}
\label{optim_prob_fading2}
\begin{split}
& \min P_{k+1} = \frac{a^2 P_{k} \bar{r}_k}{\bar{c}_k^2 P_{k} + \bar{r}_k} + \sigma_w^2 \\
& \mbox{ subject to }  \sum_{i=1}^M \alpha_{i,k}^2 \left( \frac{c_i^2 \sigma_w^2}{1-a^2} + \sigma_i^2 \right) \leq \gamma_{total}
\end{split}
\end{equation}
As we can rewrite the objective as
$$ \frac{a^2 P_k \bar{r}_k/\bar{c}_k^2}{P_k+\bar{r}_k/\bar{c}_k^2} + \sigma_w^2$$
it is clear that minimizing the objective function is equivalent to minimizing $\bar{r}_k/\bar{c}_k^2$. So at each time step we essentially solve the same problem (\ref{P2}) considered in Section \ref{min_distortion_sec}, while updating the value of $P_{k+1}$ every time. 

\subsubsection{Orthogonal access}
Recall from (\ref{Riccati_eqn2}) that in the orthogonal scheme, the Kalman filter recursion for the error covariance is:
$$P_{k+1}^o = \frac{a^2 P_k^o}{1 + P_k^o \bar{\textbf{C}}_k^{o^T} \bar{\textbf{R}}_k^{o^{-1}} \bar{\textbf{C}}_k^o} + \sigma_w^2
$$

If we wish to minimize the sum power while keeping $P_{k+1}^o \leq D$ at all time instances, the constraint becomes 
$$\bar{\textbf{C}}_k^{o^T} \bar{\textbf{R}}_k^{o^{-1}} \bar{\textbf{C}}_k^o = \sum_{i=1}^M \frac{\alpha_{i,k}^2 h_{i,k}^2 c_i^2}{\alpha_{i,k}^2 h_{i,k}^2 \sigma_i^2 + \sigma_n^2} \geq \frac{a^2 P_k^o + \sigma_w^2-D}{P_k^o (D-\sigma_w^2)}$$

If we wish to minimize $P_{k+1}^o$ at each time instance subject to a sum power constraint at all times $k$, then this is the same as maximizing 
$$\bar{\textbf{C}}_k^{o^T} \bar{\textbf{R}}_k^{o^{-1}} \bar{\textbf{C}}_k^o = \sum_{i=1}^M \frac{\alpha_{i,k}^2 h_{i,k}^2 c_i^2}{\alpha_{i,k}^2 h_{i,k}^2 \sigma_i^2 + \sigma_n^2}$$

In both cases, the resulting optimization problems which are to be solved at each time instant are variants of problems (\ref{P3}) and (\ref{P4}), and can be handled using the same techniques.

\subsubsection{Remarks}
As discussed in Section \ref{optim_interpretation}, these problems can be solved in a distributed manner, with the fusion center broadcasting some global constants that can then be used by the individual sensors to computer their optimal power allocation. 
The main issue with running these optimizations at every time step  is the cost of obtaining channel state information. If the channels don't vary too quickly one might be able to use the same values for the channel gains over a number of different time steps. However if the channels vary quickly then estimating the channels at each time step may not be feasible or practical. In this case we propose one possible alternative, which is the use of a linear estimator that depends only on the channel statistics, and which will be derived in Section \ref{no_CSI_sec}. 

\subsubsection{A dynamic programming formulation}
\label{DP_sec}
The optimization problems we have formulated in this section follow a ``greedy'' approach where we have constraints that must be satisfied at each time step, which allows us to use the same techniques as in Sections \ref{mac_optim_sec} and \ref{orth_optim_sec}. The motivation behind this follows from the monotonic properties of the solution to the Riccati equations (\ref{Riccati_eqn1}) or (\ref{Riccati_eqn2}). An alternative formulation is to consider constraints on the long term averages of the estimation error and transmission powers. For instance, instead of problem (\ref{optim_prob_fading2}), one might consider instead the infinite horizon problem:
\begin{equation*}
\begin{split}
& \min \lim_{T \rightarrow \infty} \frac{1}{T} \sum_{k=1}^T \mathbb{E}\left[P_{k+1}  \right]\\
& \mbox{ subject to } \lim_{T \rightarrow \infty} \frac{1}{T} \sum_{k=1}^T \mathbb{E}[ \sum_{i=1}^M \gamma_{i,k} ] \leq \gamma_{total}
\end{split}
\end{equation*}
where we wish to determine policies that will minimize the expected error covariance subject to the average sum power being less than a threshold $\gamma_{total}$.  
Solving such problems will require dynamic programming techniques, and would involve discretization of the optimization variables similar to \cite{GhasemiDey}, where optimal quantizers were designed for HMM state estimation over bandwidth contrained channels using a stochastic control approach. This approach is however highly computationally demanding. 
A thorough study of these problems is beyond the scope of this paper and is currently under investigation.  

\subsection{Fading channels without CSI}
\label{no_CSI_sec}
Suppose now that CSI is not available at either the sensors or fusion center, though channel statistics are available.\footnote{We note that this can also be used to model the situation where the sensors are not perfectly synchronized \cite{Gastpar_JSAC}.} The optimal filters in this case will be nonlinear and highly complex, see e.g. \cite{JafferGupta1}. An alternative is to consider the best \emph{linear} estimator in the minimum mean squared error (MMSE) sense, based on \cite{Rajasekaran}. In our notation, the situation considered in \cite{Rajasekaran} would be applicable to the model
$
x_{k+1}  =  a x_k + w_k, 
z_{k}  =  \alpha_k h_k c x_k + v_k$.
While this is not quite the same as the situations that we are considering in this paper, their techniques 
can be suitably extended. 

\subsubsection{Multi-access scheme}
\label{no_CSI_mac}
 
Since we do not have CSI we cannot do transmitter beamforming and must return to the full complex model (\ref{mac_complex_received}). We will also restrict  $\tilde{\alpha}_{i,k} = \tilde{\alpha}_{i}, \forall k$ to be time invariant. The main difference from \cite{Rajasekaran} is that the innovations is now defined as 
$$\left[\begin{array}{c} \Re[\tilde{z}_k] \\ \Im[\tilde{z}_k] \end{array} \right] - \left[\begin{array}{c} \sum_{i=1}^M \mathbb{E}[\Re[\tilde{\alpha}_i \tilde{h}_i]] c_i\\ \sum_{i=1}^M \mathbb{E}[\Im[\tilde{\alpha_i} \tilde{h}_i]] c_i \end{array} \right] \hat{x}_{k|k-1}$$
Assuming that the processes $\{\tilde{h}_{i,k}\}, i = 1, \dots, M$ are i.i.d. over time, with real and imaginary components independent of each other, and $\{\tilde{h}_{i,k}\}$ independent of $\{w_k\}$ and $\{v_{i,k}\}, i = 1, \dots, M$, the linear MMSE estimator for scalar systems can then be derived following the methods of \cite{Rajasekaran} (also see \cite{Tugnait_multiplicative}) as follows: 
\begin{equation}
\begin{split}
\label{MMSE_eqns}
\hat{x}_{k+1|k} & =  a \hat{x}_{k|k} \\
P_{k+1|k} & =  a^2 P_{k|k} \\
\hat{x}_{k+1|k+1} & =  \hat{x}_{k+1|k} + P_{k+1|k} \bar{\bar{\textbf{C}}}^T \left(\bar{\bar{\textbf{C}}} P_{k+1|k} \bar{\bar{\textbf{C}}}^T + \bar{\bar{\textbf{R}}}\right)^{-1} \left((\Re[\tilde{z}_{k+1}] , \Im[\tilde{z}_{k+1}] )^T - \bar{\bar{\textbf{C}}} \hat{x}_{k+1|k}\right) \\
P_{k+1|k+1} & =  P_{k+1|k} - P_{k+1|k}^2 \bar{\bar{\textbf{C}}}^T \left(\bar{\bar{\textbf{C}}}  P_{k+1|k} \bar{\bar{\textbf{C}}}^T + \bar{\bar{\textbf{R}}}\right)^{-1}
\end{split}
\end{equation}
where
$
\bar{\bar{\textbf{C}}}  \equiv  \left[\begin{array}{cc} \sum_{i=1}^M \mathbb{E}[\Re[\tilde{\alpha}_i \tilde{h}_i]] c_i & \sum_{i=1}^M \mathbb{E}[\Im[\tilde{\alpha_i} \tilde{h}_i]] c_i \end{array} \right]^T $ and
$$\bar{\bar{\textbf{R}}}  \equiv \left[ \begin{array}{cc} \sum_{i=1}^M \left( \textrm{Var}[\Re[\tilde{\alpha_i} \tilde{h}_i]]   \frac{c_i^2 \sigma_w^2}{1-a^2} + \mathbb{E}[\Re^2[\tilde{\alpha_i} \tilde{h}_i]] \sigma_i^2 \right)  + \sigma_n^2]  &  \sum_{i=1}^M \mathbb{E}[\Re[\tilde{\alpha_i} \tilde{h}_i]] \mathbb{E}[\Im[\tilde{\alpha_i} \tilde{h}_i]]\sigma_i^2 \\ \sum_{i=1}^M \mathbb{E}[\Re[\tilde{\alpha_i} \tilde{h}_i]] \mathbb{E}[\Im[\tilde{\alpha_i} \tilde{h}_i]]\sigma_i^2 & \sum_{i=1}^M \left( \textrm{Var}[\Im[\tilde{\alpha_i} \tilde{h}_i]]   \frac{c_i^2 \sigma_w^2}{1-a^2} + \mathbb{E}[\Im^2[\tilde{\alpha_i} \tilde{h}_i]] \sigma_i^2 \right) + \sigma_n^2 \end{array} \right]
$$ 
using the shorthand $ \Re^2[X] = (\Re[X])^2$ and $\Im^2[X] = (\Im[X])^2$.

These equations look like the Kalman filter equations but with different $\textbf{C}$ and $\textbf{R}$ matrices, so much of our previous analysis will apply.\footnote{In fact one can regard it as an ``equivalent'' linear system (with a stable dynamics and stationary noise processes) along the lines of \cite{Tugnait_multiplicative}.} For instance, since the estimator is not using the instantaneous time-varying channel gains but only the channel statistics (which are assumed to be constant), there \emph{will} be a steady state error covariance given by 
$$P_\infty = \frac{(a^2-1) +  \sigma_w^2 S + \sqrt{(a^2-1 +  \sigma_w^2 S)^2 + 4  \sigma_w^2 S }}{2 S}$$
with $S \equiv \bar{\bar{\textbf{C}}}^T \bar{\bar{\textbf{R}}}^{-1}\bar{\bar{\textbf{C}}}$. Note that for circularly symmetric fading channels e.g. Rayleigh, we have $\bar{\bar{\textbf{C}}} = [\begin{array}{cc} 0 & 0 \end{array}]$, and estimates obtained using this estimator will not be useful.\footnote{Other work where there are difficulties with circularly symmetric fading include \cite{MergenTong,Gastpar_JSAC,LiuElGamalSayeed}. A possible scheme for estimation of i.i.d. processes and zero-mean channels which can achieve a $1/\log M$ scaling has been proposed in \cite{LiuElGamalSayeed}.}
Thus we will now restrict ourselves to non-zero mean fading processes. 
Motivated by transmitter beamforming in the case with CSI, let us use amplification factors of the form
$$\tilde{\alpha}_i = \alpha_i \frac{(\mathbb{E}[\tilde{h}_i])^*}{|\mathbb{E}[\tilde{h}_i]|}$$
with $\alpha_i \in \mathbb{R}$. Then $S$ simplifies to 
$$S= \frac{\left(\sum_{i=1}^M \mathbb{E}[\Re[\tilde{\alpha}_i \tilde{h}_i]] c_i\right)^2}{\sum_{i=1}^M \left( \textrm{Var}[\Re[\tilde{\alpha_i} \tilde{h}_i]]  c_i^2 \frac{\sigma_w^2}{1-a^2} + \mathbb{E}[\Re^2[\tilde{\alpha_i} \tilde{h}_i]] \sigma_i^2 \right)  + \sigma_n^2 }$$
where we can find 
\begin{equation}
\label{no_CSI_statistics}
\begin{split}
\mathbb{E}[\Re[\tilde{\alpha}_i \tilde{h}_i]] & =  \alpha_i |\mathbb{E}[\tilde{h}_i]| \\
\textrm{Var}[\Re[\tilde{\alpha_i} \tilde{h}_i]] & =  \frac{\alpha_i^2}{|\mathbb{E}[\tilde{h}_i]|^2} \left(\mathbb{E}^2[\Re \tilde{h}_i]\textrm{Var}[\Re \tilde{h}_i] + \mathbb{E}^2[\Im \tilde{h}_i]\textrm{Var}[\Im \tilde{h}_i]\right) \\
\mathbb{E}[\Re^2[\tilde{\alpha_i} \tilde{h}_i]] & =  \frac{\alpha_i^2}{|\mathbb{E}[\tilde{h}_i]|^2} \left( \mathbb{E}^2[\Re \tilde{h}_i] \mathbb{E}[\Re^2 \tilde{h}_i] + 2 \mathbb{E}^2[\Re \tilde{h}_i] \mathbb{E}^2[\Im \tilde{h}_i] + \mathbb{E}^2[\Im \tilde{h}_i] \mathbb{E}[\Im^2 \tilde{h}_i] \right)
\end{split}
\end{equation}
using the shorthand $\mathbb{E}^2 [X] = (\mathbb{E}[X])^2, \Re^2[X] = (\Re[X])^2$ and $\Im^2[X] = (\Im[X])^2$. 
If the real and imaginary parts are identically distributed, we have the further simplifications $\textrm{Var}[\Re[\tilde{\alpha_i} \tilde{h}_i]] = \alpha_i^2 \textrm{Var}[\Re \tilde{h}_i]$ and $\mathbb{E}[\Re^2[\tilde{\alpha_i} \tilde{h}_i]] = \alpha_i^2 \left( \mathbb{E}[\Re^2 \tilde{h}_i] + \mathbb{E}^2[\Re \tilde{h}_i] \right)$.

Power allocation using this sub-optimal estimator can then be developed, and the resulting optimization problems (which are omitted for brevity) will be variants of problems (\ref{P1}) and (\ref{P2}). We note however that the optimization problems will only need to be run \emph{once} since $\bar{\bar{\textbf{C}}}$ and $\bar{\bar{\textbf{R}}}$ are  time-invariant quantities, rather than at each time instance as in the case with CSI. 

Since we have a steady state error covariance using this estimator, asymptotic behaviour can also be analyzed by using the  techniques in Sections \ref{asymptotic_sec}. The details are omitted for brevity.

\subsubsection{Orthogonal access scheme}
\label{no_CSI_orth}
For orthogonal access and no CSI, the equations for the linear MMSE can also be similarly derived and will be of the form (\ref{MMSE_eqns}), substituting $\bar{\bar{\textbf{C}}}^o$ in place of $\bar{\bar{\textbf{C}}}$, $\bar{\bar{\textbf{R}}}^o$ in place of $\bar{\bar{\textbf{R}}}$, etc. We have
$\bar{\bar{\textbf{C}}}^o \equiv \left[\begin{array}{ccccc} \mathbb{E}[\Re[\tilde{\alpha}_1 \tilde{h}_1]] c_1 & \mathbb{E}[\Im[\tilde{\alpha}_1 \tilde{h}_1]] c_1 & \dots & \mathbb{E}[\Re[\tilde{\alpha}_M \tilde{h}_M]] c_M & \mathbb{E}[\Im[\tilde{\alpha}_M \tilde{h}_M]] c_M \end{array} \right]^T$ and
$$ \bar{\bar{\textbf{R}}}^o \equiv \left[\begin{array}{ccc} \bar{\bar{\textbf{R}}}_{11}^o &  \dots & 0 \\  \vdots &  \ddots & \vdots \\ 0 &  \dots & \bar{\bar{\textbf{R}}}_{MM}^o \end{array} \right]$$
with each $\bar{\bar{\textbf{R}}}_{ii}^o$ being a block matrix 
$$\bar{\bar{\textbf{R}}}_{ii}^o  \equiv \left[ \begin{array}{cc}  \textrm{Var}[\Re[\tilde{\alpha_i} \tilde{h}_i]]  c_i^2 \frac{\sigma_w^2}{1-a^2} + \mathbb{E}[\Re^2[\tilde{\alpha_i} \tilde{h}_i]] \sigma_i^2   + \sigma_n^2  & \mathbb{E}[\Re[\tilde{\alpha_i} \tilde{h}_i]] \mathbb{E}[\Im[\tilde{\alpha_i} \tilde{h}_i]]\sigma_i^2 \\ \mathbb{E}[\Re[\tilde{\alpha_i} \tilde{h}_i]] \mathbb{E}[\Im[\tilde{\alpha_i} \tilde{h}_i]]\sigma_i^2 &  \textrm{Var}[\Im[\tilde{\alpha_i} \tilde{h}_i]]  c_i^2 \frac{\sigma_w^2}{1-a^2} + \mathbb{E}[\Im^2[\tilde{\alpha_i} \tilde{h}_i]] \sigma_i^2  + \sigma_n^2 \end{array} \right]
$$
There will be a steady state error covariance given by 
$$P_\infty^o = \frac{(a^2-1) +  \sigma_w^2 S^o + \sqrt{(a^2-1 +  \sigma_w^2 S^o)^2 + 4  \sigma_w^2 S^o }}{2 S^o}$$
with $S^o=\bar{\bar{\textbf{C}}}^{o^T} \bar{\bar{\textbf{R}}}^{o^{-1}}\bar{\bar{\textbf{C}}}^o$. 
If we choose
$\tilde{\alpha}_i = \alpha_i \frac{(\mathbb{E}[\tilde{h}_i])^*}{|\mathbb{E}[\tilde{h}_i]|}$
then $S^o$ can be shown to be
$$S^o = \sum_{i=1}^M \frac{\left( \mathbb{E}[\Re[\tilde{\alpha}_i \tilde{h}_i]] c_i\right)^2}{ \left( \textrm{Var}[\Re[\tilde{\alpha_i} \tilde{h}_i]]  c_i^2 \frac{\sigma_w^2}{1-a^2} + \mathbb{E}[\Re^2[\tilde{\alpha_i} \tilde{h}_i]] \sigma_i^2 \right)  + \sigma_n^2 }$$
where we also refer to (\ref{no_CSI_statistics}) for further simplifications of these quantities. 

Asymptotic behaviour and optimal power allocation can also be analyzed using the  techniques in Sections \ref{asymptotic_sec} and \ref{orth_optim_sec} respectively, and the details are omitted for brevity.

\section{Numerical studies}
\label{numerical_sec}

\subsection{Static channels}
\label{mac_static_numerical}
\begin{figure}[tbp]
\centering
\includegraphics[width=14.0cm]{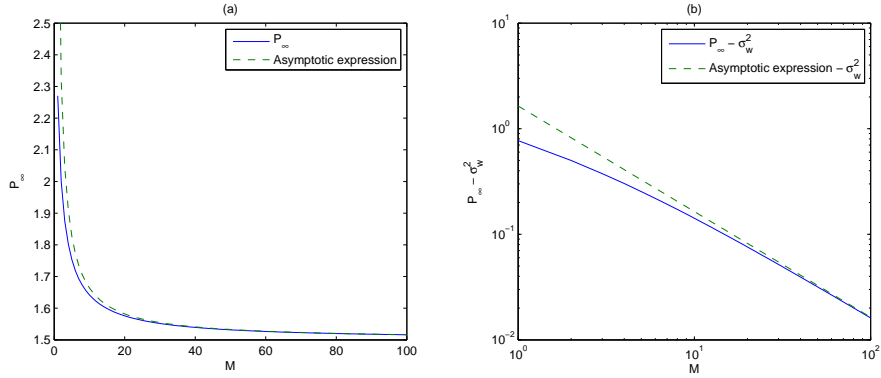}
\caption{Comparison between $P_\infty$ and asymptotic expression: Multi-access scheme with $\alpha_i = 1/\sqrt{M}$}
\label{asymptotic_plot1}
\end{figure}

First we show some plots for the asymptotic results of Section \ref{asymptotic_sec}. In Fig. \ref{asymptotic_plot1} (a) we plot $P_\infty$ vs $M$ in the multi-access scheme for the symmetric situation with $\alpha_i = 1/\sqrt{M}$, and $a=0.8, \sigma_w^2=1.5, \sigma_n^2=1,c=1,\sigma_v^2=1,h=0.8$. We compare this with the asymptotic expression 
$\sigma_w^2 + \frac{a^2 (\sigma_v^2+\sigma_n^2/h^2)}{c^2}\frac{1}{M}$ from (\ref{asympt_expr_2}).
Fig. \ref{asymptotic_plot1} (b) plots the difference between $P_\infty-\sigma_w^2$, and compares this with the term $\frac{a^2 (\sigma_v^2+\sigma_n^2/h^2)}{c^2}\frac{1}{M}$. We can see that  $P_\infty$ is well approximated by the asymptotic expression even for 20-30 sensors. 

\begin{figure}[tbp]
\centering
\includegraphics[width=14.0cm]{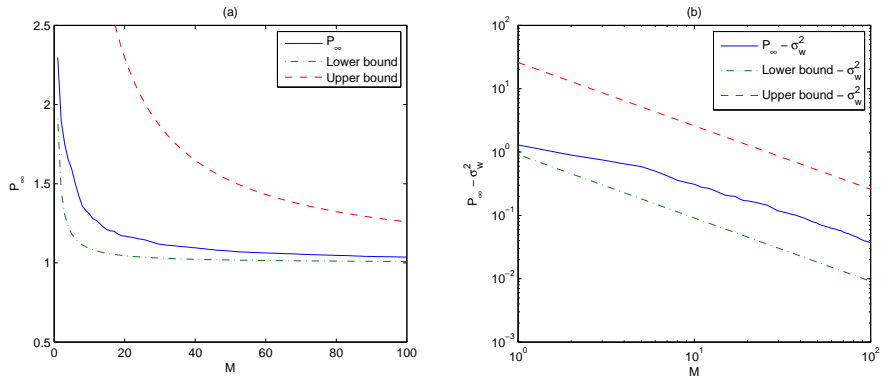}
\caption{$P_\infty$ with general parameters and bounds: Multi-access scheme with $\alpha_i = 1/\sqrt{M}$}
\label{asymptotic_plot2}
\end{figure}

In Fig. \ref{asymptotic_plot2} we plot $P_\infty$ vs $M$ in the multi-access scheme with $\alpha_i = 1/\sqrt{M}, a=0.9, \sigma_w^2=1, \sigma_n^2=1$ and values for $c_i, \sigma_i^2, h_i$ chosen from the range $0.5 \leq C_i \leq 1, 0.5 \leq R_i \leq 1, 0.5\leq h_i \leq 1$. We also plot the (asymptotic) lower and upper bounds (\ref{asympt_bounds2}) from the proof of Lemma \ref{asympt_lemma2}, $\sigma_w^2 + \frac{a^2 (h_{min}^2 \sigma_{min}^2+\sigma_n^2)}{h_{max}^2 c_{max}^2}\frac{1}{M}$ and $\sigma_w^2 + \frac{a^2 (h_{max}^2 \sigma_{max}^2+\sigma_n^2)}{h_{min}^2 c_{min}^2}\frac{1}{M}$. It can be seen that $P_\infty$ does indeed lie between the two bounds, both of which converge to $\sigma_w^2$ at the rate $1/M$. 

\begin{figure}[tbp]
\centering
\includegraphics[width=14.0cm]{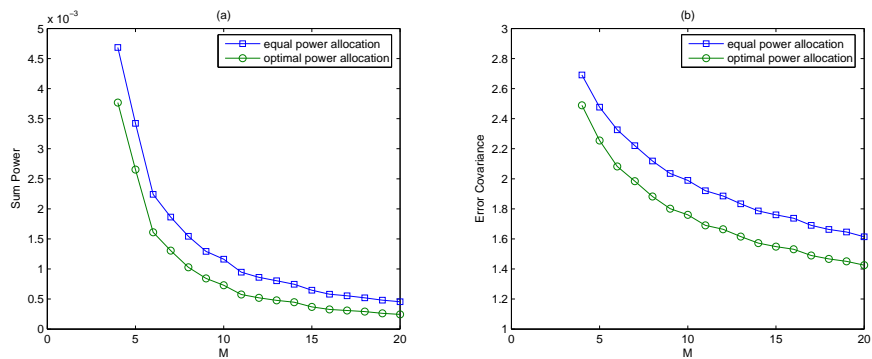}
\caption{Multi-access. Comparison between optimal and equal power allocation schemes, with (a) an error covariance constraint and (b) a sum power constraint}
\label{mac_static}
\end{figure}

\begin{figure}[tbp]
\centering
\includegraphics[width=14.0cm]{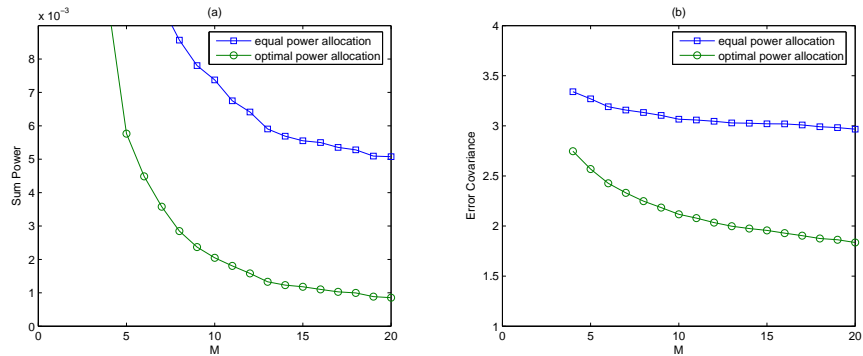}
\caption{Orthogonal access. Comparison between optimal and equal power allocation schemes, with (a) an error covariance constraint and (b) a sum power constraint}
\label{orth_static}
\end{figure}

Next we look at the numerical results for optimal power allocation. In Fig. \ref{mac_static} we compare between using optimal power allocation and equal power allocation for the multi-access scheme. We use $a=0.9, \sigma_n^2=10^{-9}, \sigma_w^2=1, c_i=1, \forall i$. The sensor noise variances $ \sigma_i^2$ are drawn from a $\chi^2(1)$ distribution to model the differences in sensor measurement quality.  
The channel gains $h_i$ are modelled as $d_i^{-2}$, with $d_i$ representing the distance of sensor $i$ to the fusion center. We use distances uniformly drawn between 20m and 100m. In Fig. \ref{mac_static}(a) we keep $D = 2$, while in Fig. \ref{mac_static}(b) we keep $\gamma_{total} = 10^{-3}$. Each of the data points represent the average over 1000 realisations of the sensor parameters (i.e. $c_i, \sigma_i^2, d_i$). 
In Fig. \ref{orth_static} the comparison using the same parameters and parameter distributions is shown for the orthogonal scheme.
What can be observed is that as the number of sensors $M$ increases there is a general trend downwards for both graphs, though optimal power allocation seems to provide more benefits in the orthogonal access scheme than the multi-access scheme. 

\subsection{Fading channels}
\label{mac_fading_numerical}
\begin{figure}[tbp]
\centering
\includegraphics[width=14.0cm]{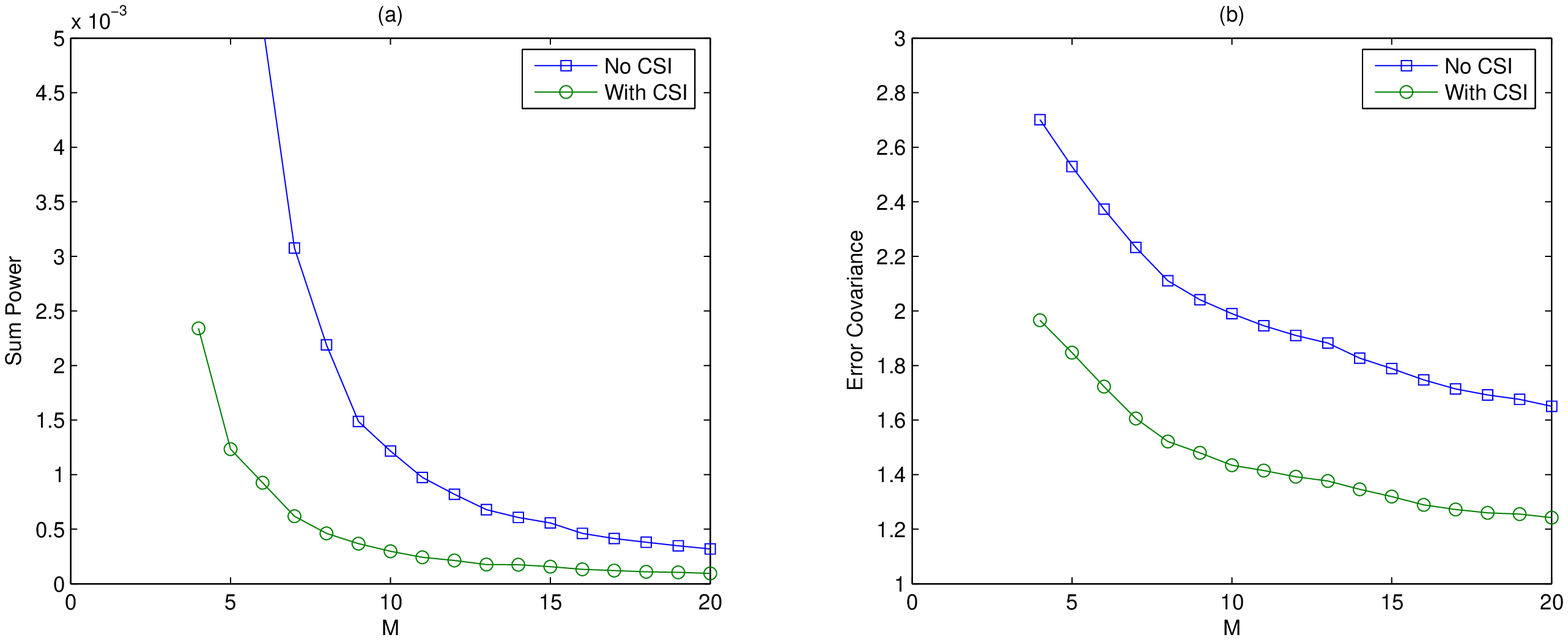}
\caption{Multi-access. Comparison between the full CSI and no CSI situations, with (a) an error covariance constraint and (b) a sum power constraint}
\label{mac_fading}
\end{figure}

\begin{figure}[tbp]
\centering
\includegraphics[width=14.0cm]{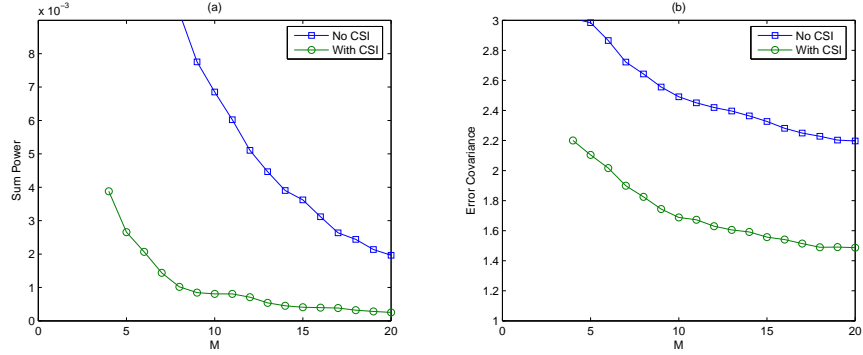}
\caption{Orthogonal access. Comparison between the full CSI and no CSI situations, with (a) an error covariance constraint and (b) a sum power constraint}
\label{orth_fading}
\end{figure}

In Fig. \ref{mac_fading} we compare between the full CSI and no CSI situations for the multi-access scheme, using $a=0.9, \sigma_n^2=10^{-9}, \sigma_w^2=1, c_i=1, \forall i$, and $ \sigma_i^2$ drawn from a $\chi^2(1)$ distribution. The complex channel gains $\tilde{h}_{i,k}$'s are chosen to be Rician distributed with distance dependence. Specifically, the real and imaginary parts of $\tilde{h}_{i,k}$ are chosen to be distributed as $d_i^{-2} \times N(\mu_i,1)$, with $d_i$ uniform between 20 and 100, and $\mu_i$ uniform between 1/2 and 1. In Fig. \ref{mac_fading}(a) we keep $D = 2$, and in Fig. \ref{mac_fading}(b) we keep $\gamma_{total} = 10^{-3}$. In the full CSI case the values are averaged over 1000 time steps for each set of sensor parameters (i.e. $c_i, \sigma_i^2, d_i, \mu_i$), and in the no CSI case they are the steady state values using the linear MMSE estimator (\ref{MMSE_eqns}). The results are then repeated and further averaged over 100 realisations of the sensor parameters. 
In Fig. \ref{orth_fading} we make the same comparison for the orthogonal scheme. 
We can  see in Fig. \ref{mac_fading} that for the multi-access scheme the performance loss in the case without CSI is not too great when compared to the case with full CSI. Thus even if one has full CSI, but doesn't want to perform power allocation at every time step, using the linear MMSE estimator (\ref{MMSE_eqns}) instead could be an attractive alternative. On the other hand, for the orthogonal scheme in Fig. \ref{orth_fading} there is a more significant performance loss in the situation with no CSI.

\section{Extension to vector states and MIMO}
\label{extensions_sec}
In Section \ref{vector_sec} we formulate a possible extension of our work to vector state linear systems. We outline some of the differences and difficulties that will be encountered when compared with the scalar case. In Section \ref{MIMO_sec} we consider a situation similar to a MIMO system, where the fusion center has multiple receive antennas (and each sensor operating with a single transmit antenna), and we show how they can be written as an equivalent vector linear system. 
 
\subsection{Vector states}
\label{vector_sec}
We consider a general vector model
$$ \textbf{x}_{k+1} = \textbf{A} \textbf{x}_k + \textbf{w}_k$$
with $\textbf{x} \in \mathbb{R}^n$, $\textbf{A} \in \mathbb{R}^{n \times n}$, and $\textbf{w}_k \in \mathbb{R}^n$ being Gaussian with zero-mean and covariance matrix \textbf{Q}. For a stable system all the eigenvalues of the matrix $\textbf{A}$ will have magnitude less than 1. 
The $M$ sensors each observe
$$ \textbf{y}_{i,k} = \textbf{C}_i \textbf{x}_k + \textbf{v}_{i,k} , i = 1, \dots , M$$
with $\textbf{y}_{i,k} \in \mathbb{R}^m$, $\textbf{C}_i \in \mathbb{R}^{m\times n}$, and $\textbf{v}_{i,k} \in \mathbb{R}^m$ being Gaussian with zero-mean and covariance matrix $\textbf{R}_i$. 
We assume that each of the individual components of the measurement vectors $\textbf{y}_{i,k}$ are amplified and forwarded to a fusion center via separate orthogonal channels.\footnote{Another possibility is to apply compression on the measured signal \cite{SchizasGiannakisLuo,Xiao_coherent_TSP}, so that the dimensionality of the signal that the sensor transmits is smaller than the dimension of the measurement vector, but for simplicity we will not consider this here.} We will consider real channel gains for simplicity. 

In the multi-access scheme the fusion center then receives
$$
\textbf{z}_k  =  \sum_{i=1}^M \textbf{H}_{i,k} \boldsymbol\alpha_{i,k} \textbf{y}_{i,k} + \textbf{n}_k
$$
where $\boldsymbol\alpha_{i,k} \in \mathbb{R}^{m \times m}$ is a matrix of amplification factors, $\textbf{H}_{i,k} \in \mathbb{R}^{m \times m}$ a matrix of channel gains, and $\textbf{n}_k \in \mathbb{R}^m$ is Gaussian with zero-mean and covariance matrix $\textbf{N}$. 
We can express the situation as 
$$\textbf{x}_{k+1} = \textbf{A} \textbf{x}_k + \textbf{w}_k, \phantom{aa} \textbf{z}_k = \bar{\textbf{C}}_k \textbf{x}_k + \bar{\textbf{v}}_k$$
where $\bar{\textbf{C}}_k \equiv \sum_{i=1}^M \textbf{H}_{i,k} \boldsymbol\alpha_{i,k} \textbf{C}_i $, $\bar{\textbf{v}}_k \equiv \sum_{i=1}^M \textbf{H}_{i,k} \boldsymbol\alpha_{i,k} \textbf{v}_{i,k} + \textbf{n}_k$, with $\bar{\textbf{v}}_k$ having covariance matrix $\bar{\textbf{R}}_k \equiv \sum_{i=1}^M \textbf{H}_{i,k} \boldsymbol\alpha_{i,k} \textbf{R}_i \boldsymbol\alpha_{i,k}^T \textbf{H}_{i,k}^T + \textbf{N}$. 
The error covariance updates as follows:
$$\textbf{P}_{k+1} = \textbf{A} \textbf{P}_k \textbf{A}^T - \textbf{A} \textbf{P}_k \bar{\textbf{C}}_k^T (\bar{\textbf{C}}_k \textbf{P}_k \bar{\textbf{C}}_k^T + \bar{\textbf{R}}_k)^{-1} \bar{\textbf{C}}_k \textbf{P}_k \textbf{A}^T + \textbf{Q}$$
The transmit power of sensor $i$ at time $k$ is  
\begin{eqnarray*}
\boldsymbol\gamma_{i,k} & =&\textrm{Tr}(\boldsymbol\alpha_{i,k}  \mathbb{E}[\textbf{y}_k \textbf{y}_k^T]  \boldsymbol\alpha_{i,k}^T) \\
& = & \textrm{Tr} (\boldsymbol\alpha_{i,k} (\textbf{C}_i \mathbb{E}[\textbf{x}_k \textbf{x}_k^T] \textbf{C}_i^T + \textbf{R}_i) \boldsymbol\alpha_{i,k}^T)
\end{eqnarray*}
where $\textrm{Tr}(\bullet)$ denotes the trace, and $\mathbb{E}[\textbf{x}_k \textbf{x}_k^T]$ satisfies (see \cite[p.71]{AndersonMoore})
$$ \mathbb{E}[\textbf{x}_k \textbf{x}_k^T] - \textbf{A} \mathbb{E}[\textbf{x}_k \textbf{x}_k^T] \textbf{A}^T = \textbf{Q}$$
In the static channel case, the steady state error covariance $\textbf{P}_\infty$ satisfies 
$$\textbf{P}_{\infty} = \textbf{A} \textbf{P}_\infty \textbf{A}^T - \textbf{A} \textbf{P}_\infty \bar{\textbf{C}}^T (\bar{\textbf{C}} \textbf{P}_\infty \bar{\textbf{C}}^T + \bar{\textbf{R}})^{-1} \bar{\textbf{C}} \textbf{P}_\infty \textbf{A}^T + \textbf{Q}$$
However, unlike the scalar case where the closed form expression (\ref{P_inf_scalar}) exists, in the vector case no such formula for $\textbf{P}_{\infty}$ is available, and thus asymptotic analysis is difficult to develop. 
For time-varying channels, we can pose similar optimization problems as considered in Section \ref{optim_sec}. For instance, minimization of the error covariance subject to a sum power constraint can be written as:
\begin{equation}
\label{P5}
\begin{split}
& \min_{\boldsymbol\alpha_{1,k},\dots,\boldsymbol\alpha_{M,k}} \textrm{Tr}(\textbf{P}_{k+1})  \\
& \mbox{ subject to }  \sum_{i=1}^M (\boldsymbol\alpha_{i,k} (\textbf{C}_i \mathbb{E}[\textbf{x}_k \textbf{x}_k^T] \textbf{C}_i^T + \textbf{R}_i) \boldsymbol\alpha_{i,k}^T) \leq \gamma_{total}
\end{split}
\end{equation} 
This problem is non-convex, and unlike the scalar case does not appear to be able to be reformulated into a convex problem. Similar problems have been considered previously in the context of parameter estimation, and sub-optimal solutions were presented using techniques such as deriving bounds on the error covariance \cite{FangLi}, and convex relaxation techniques \cite{Xiao_coherent_TSP}. 
 
In the orthogonal access scheme the fusion center receives
$$ \textbf{z}_{i,k} = \textbf{H}_{i,k} \boldsymbol\alpha_{i,k} \textbf{y}_{i,k} + \textbf{n}_{i,k}, i = 1,\dots, M$$
We can express the situation as
$$\textbf{x}_{k+1} = \textbf{A} \textbf{x}_k + \textbf{w}_k, \phantom{aa} \textbf{z}_k^o = \bar{\textbf{C}}_k^o \textbf{x}_k + \bar{\textbf{v}}_k^o$$
by defining 
$$\textbf{z}_k^o \equiv \left[\begin{array}{c} \textbf{z}_{1,k} \\ \vdots \\ \textbf{z}_{M,k} \end{array} \right], \bar{\textbf{C}}_k^o \equiv \left[\begin{array}{c} \textbf{H}_{1,k} \boldsymbol\alpha_{1,k}  \textbf{C}_1 \\ \vdots \\ \textbf{H}_{M,k} \boldsymbol\alpha_{M,k}  \textbf{C}_M \end{array} \right], \bar{\textbf{v}}_k^o \equiv \left[\begin{array}{c} \textbf{H}_{1,k} \boldsymbol\alpha_{1,k}  \textbf{v}_{1,k}+\textbf{n}_{1,k} \\ \vdots \\ \textbf{H}_{M,k} \boldsymbol\alpha_{M,k}  \textbf{v}_{M,k} + \textbf{n}_{M,k} \end{array} \right]$$
with the covariance of $\bar{\textbf{v}}_k^o$ being
$$\bar{\textbf{R}}_k^o \equiv \left[\begin{array}{cccc} \textbf{H}_{1,k} \boldsymbol\alpha_{1,k} \textbf{R}_1 \boldsymbol\alpha_{1,k}^T \textbf{H}_{1,k}^T + \textbf{N} & 0 & \dots & 0 \\ 0 & \textbf{H}_{2,k} \boldsymbol\alpha_{2,k} \textbf{R}_2 \boldsymbol\alpha_{2,k}^T \textbf{H}_{2,k}^T + \textbf{N}  & \dots & 0 \\ \vdots & \vdots & \ddots & \vdots \\ 0 & 0 & \dots & \textbf{H}_{M,k} \boldsymbol\alpha_{M,k} \textbf{R}_M \boldsymbol\alpha_{M,k}^T \textbf{H}_{M,k}^T + \textbf{N} \end{array} \right]$$
The error covariance updates as follows:
$$\textbf{P}_{k+1}^o = \textbf{A} \textbf{P}_k^o \textbf{A}^T - \textbf{A} \textbf{P}_k^o \bar{\textbf{C}}_k^{o^T} (\bar{\textbf{C}}^o_k \textbf{P}_k^o \bar{\textbf{C}}_k^{o^T} + \bar{\textbf{R}}_k^o)^{-1} \bar{\textbf{C}}_k^o \textbf{P}_k^o \textbf{A}^T + \textbf{Q}$$
The term $\bar{\textbf{C}}_k^{o^T} (\bar{\textbf{C}}^o_k \textbf{P}_k^o \bar{\textbf{C}}_k^{o^T} + \bar{\textbf{R}}_k^o)^{-1} \bar{\textbf{C}}_k^o$ can be rewritten using the matrix inversion lemma as 
$$\bar{\textbf{C}}_k^{o^T} (\bar{\textbf{C}}^o_k \textbf{P}_k^o \bar{\textbf{C}}_k^{o^T} + \bar{\textbf{R}}_k^o)^{-1} \bar{\textbf{C}}_k^o = \textbf{C}_k^{o^T} \bar{\textbf{R}}_k^{o^{-1}} \bar{\textbf{C}}_k^{o^T} - \textbf{C}_k^{o^T} \bar{\textbf{R}}_k^{o^{-1}} \bar{\textbf{C}}_k^{o^T} (\textbf{P}_k^{o^{-1}} + \textbf{C}_k^{o^T} \bar{\textbf{R}}_k^{o^{-1}} \bar{\textbf{C}}_k^{o^T})^{-1} \textbf{C}_k^{o^T} \bar{\textbf{R}}_k^{o^{-1}} \bar{\textbf{C}}_k^{o^T}$$
where we have the simplification 
$$\textbf{C}_k^{o^T} \bar{\textbf{R}}_k^{o^{-1}} \bar{\textbf{C}}_k^{o^T} = \sum_{i=1}^M (\textbf{H}_{i,k} \boldsymbol\alpha_{i,k} \textbf{C}_i)^T (\textbf{H}_{i,k} \boldsymbol\alpha_{i,k} \textbf{R}_i \boldsymbol\alpha_{i,k}^T \textbf{H}_{i,k}^T + \textbf{N})^{-1} (\textbf{H}_{i,k} \boldsymbol\alpha_{i,k} \textbf{C}_i)$$
Minimization of the error covariance subject to a sum power constraint can be written as:
\begin{equation}
\label{P6}
\begin{split}
& \min_{\boldsymbol\alpha_{1,k},\dots,\boldsymbol\alpha_{M,k}} \textrm{Tr}(\textbf{P}_{k+1}^o)  \\
& \mbox{ subject to }  \sum_{i=1}^M (\boldsymbol\alpha_{i,k} (\textbf{C}_i \mathbb{E}[\textbf{x}_k \textbf{x}_k^T] \textbf{C}_i^T + \textbf{R}_i) \boldsymbol\alpha_{i,k}^T) \leq \gamma_{total}
\end{split}
\end{equation} 
This problem is non-convex and also does not appear to be able to be reformulated into a convex problem. In the context of parameter estimation with sensors communicating to a fusion center via orthogonal channels, a similar problem was considered in \cite{LuoGiannakisZhang}, and was in fact shown to be NP-hard, although sub-optimal methods for solving that problem were later studied in \cite{SchizasGiannakisLuo}. 

As the techniques involved are quite different from what has currently been presented, a comprehensive study of optimization problems such as (\ref{P5}) and (\ref{P6}) is beyond the scope of this paper and will be studied elsewhere. 

\subsection{MIMO situation}
\label{MIMO_sec}
One could also consider a situation resembling the MIMO systems in wireless communications, with the different sensors (each with a single transmit antenna) representing the multiple transmitters, and multiple receive antennas at the fusion center. It turns out that these situations can be expressed as equivalent vector linear systems. We will show how this is done for a simple case. 
Consider the vector state, scalar measurement system
$$
\textbf{x}_{k+1}  = \textbf{A} \textbf{x}_k+\textbf{w}_k, \phantom{aa} 
y_{i,k}  =  \textbf{c}_i \textbf{x}_k + v_{i,k}, i=1,\dots,M
$$
where $\textbf{c}_i, \forall i$ are $1 \times n$ vectors. 
We will look at the orthogonal access scheme, but now with $L$ receive antennas at the fusion center.
The fusion center then receives from each sensor
$$\textbf{z}_{i,k} = [h_{i,k}^1 \alpha_{i,k} y_{i,k} + n_{i,k}^1, \dots, h_{i,k}^L \alpha_{i,k} y_{i,k} + n_{i,k}^L]^T, i=1, \dots, M$$
where $h_{i,k}^j$ is the channel gain from the $i$-th sensor to the $j$-th antenna. 
Defining 
$$\textbf{z}_k \equiv \left[\begin{array}{c} \textbf{z}_{1,k} \\ \vdots \\ \textbf{z}_{M,k} \end{array} \right], \bar{\textbf{C}}_k \equiv \left[h_{1,k}^1 \alpha_{1,k} \textbf{c}_1^T| \dots| h_{1,k}^L \alpha_{1,k} \textbf{c}_1^T| \dots \dots| h_{M,k}^1 \alpha_{M,k} \textbf{c}_M^T| \dots| h_{M,k}^L \alpha_{M,k} \textbf{c}_M^T\right]^T $$
$$\bar{\textbf{v}}_k \equiv  [h_{1,k}^1 \alpha_{1,k} v_{1,k}+ n_{1,k}^1,\dots, h_{1,k}^L \alpha_{1,k} v_{1,k}+ n_{1,k}^L, \dots, \dots, h_{M,k}^1 \alpha_{M,k} v_{m,k}+ n_{M,k}^1,\dots, h_{M,k}^L \alpha_{m,k} v_{m,k}+ n_{M,k}^L]^T $$
we may then write the situation as the vector system:
$$
\textbf{x}_{k+1}  = \textbf{A} \textbf{x}_k+\textbf{w}_k, \phantom{aa}
\textbf{z}_k   =  \bar{\textbf{C}}_k \textbf{x}_k + \bar{\textbf{v}}_k$$
Other variations of the MIMO setup, e.g. vector sensor measurements, can be similarly transformed into equivalent vector linear systems. Note that for scalar state and 
scalar measurements per sensor, one could use similar techniques to Section II-B for  problem formulation and those of Sections IV-B and IV-D for the optimal power allocation results.
However, as described in Section \ref{vector_sec}, difficulties in analyzing general vector systems will still remain.

\section{Conclusion}
This paper has investigated the use of analog forwarding in the distributed estimation of stable scalar linear systems. We have shown a $1/M$ scaling behaviour of the error covariance in a number of different situations, and formulated and solved some optimal power allocation problems for both static and fading channels. 
We have also outlined extensions to vector linear systems and MIMO systems. Further study of these extensions and related problems will form the topics of future investigations.

\begin{appendix}

\subsection{Proof of Lemma \ref{lemma1}}
\label{appendix_P_inf}
Rewrite (\ref{P_inf_alternate}) as
$$P_\infty = \frac{(a^2-1)}{2} \frac{1}{S} + \frac{\sigma_w^2}{2} + \sqrt{\frac{(a^2-1)^2}{4} \frac{1}{S^2} + \frac{(a^2+1)\sigma_w^2}{2}  \frac{1}{S} + \frac{\sigma_w^4}{4}}$$
Taking the derivative with respect to $S$ we get
$$\frac{dP_\infty}{dS} = -\frac{a^2-1}{2} \frac{1}{S^2} - \frac{(a^2-1)^2 \frac{1}{S^3} + (a^2+1)\sigma_w^2 \frac{1}{S^2}}{4 \sqrt{\frac{(a^2-1)^2}{4} \frac{1}{S^2} + \frac{(a^2+1)\sigma_w^2}{2} \frac{1}{S}   + \frac{\sigma_w^4}{4}}}$$
To show that $\frac{dP_\infty}{dS} \leq 0 $, it is sufficient to show that 
$$ \left(\frac{(a^2-1)^2 \frac{1}{S^3} + (a^2+1)\sigma_w^2 \frac{1}{S^2}}{4 \sqrt{\frac{(a^2-1)^2}{4} \frac{1}{S^2} + \frac{(a^2+1)\sigma_w^2}{2} \frac{1}{S}   + \frac{\sigma_w^4}{4}}} \right)^2 \geq \left(\frac{a^2-1}{2} \frac{1}{S^2} \right)^2$$
Expanding and rearranging, this is equivalent to 
\begin{equation*}
\begin{split} (a^2-1)^4 & \frac{1}{S^6}  + 2 (a^2-1)^2 (a^2+1) \sigma_w^2 \frac{1}{S^5} + (a^2+1)^2 \sigma_w^4 \frac{1}{S^4} \\ & \geq (a^2-1)^4 \frac{1}{S^6}  + 2 (a^2-1)^2 (a^2+1)  \sigma_w^2 \frac{1}{S^5} + (a^2-1)^2 \sigma_w^4 \frac{1}{S^4} 
\end{split}
\end{equation*}
or $ (a^2+1)^2 \sigma_w^4 \geq (a^2-1)^2 \sigma_w^4 $,
which is certainly true. 

\subsection{Proof of Lemma \ref{asympt_lemma1}}
We first substitute the simplified expressions for $\bar{c}$ and $\bar{r}$ into (\ref{P_inf_scalar}):
\begin{equation*}
\begin{split}
& P_\infty  = \\ & \frac{(a^2-1)(M h^2 \sigma_v^2 + \sigma_n^2)+M^2 h^2 c^2 \sigma_w^2 + \sqrt{((a^2-1)(M h^2 \sigma_v^2 + \sigma_n^2)+M^2 h^2 c^2 \sigma_w^2)^2+4M^2 h^2 c^2 \sigma_w^2(M h^2 \sigma_v^2 + \sigma_n^2)}}{2M^2 h^2 c^2}
\end{split}
\end{equation*}
Regarded as a function of $M$, we are interested in the behaviour of $P_\infty$ as $M \rightarrow \infty$. Now
\begin{equation}
\label{asymptotic_calc}
\begin{split}
&  \sqrt{((a^2-1)(M h^2 \sigma_v^2+\sigma_n^2)+M^2 h^2 c^2 \sigma_w^2)^2+4M^2 h^2 c^2 \sigma_w^2(M h^2 \sigma_v^2+\sigma_n^2)} \\
& =  \left(  h^4 c^4 \sigma_w^4 M^4 + 2(a^2-1)\sigma_v^2 h^4 c^2 \sigma_w^2 M^3 + 4 h^4 c^2 \sigma_w^2 \sigma_v^2 M^3 + O(M^2) \right)^{1/2}\\
& =   h^2 c^2 \sigma_w^2 M^2\left( 1+ \frac{2(a^2+1)\sigma_v^2}{c^2 \sigma_w^2 M} + O\left(\frac{1}{M^2}\right)  \right)^{1/2} \\
& =   h^2 c^2 \sigma_w^2 M^2\left( 1 + \frac{1}{2} \frac{2(a^2+1)\sigma_v^2}{c^2 \sigma_w^2 M} + O\left(\frac{1}{M^2}\right) \right)\\
& =   h^2 c^2 \sigma_w^2 M^2+ (a^2 + 1) h^2 \sigma_v^2 M + O(1)
\end{split}
\end{equation}
where we have used the expansion
$(1+x)^{1/2} = 1 + x/2 + O(x^2)$ for $|x| < 1$ \cite[p.15]{AbramowitzStegun}, which is valid when $M$ is sufficiently large. 
Hence
$$
P_\infty = \sigma_w^2 + \frac{a^2 \sigma_v^2}{c^2}\frac{1}{M} + O\left(\frac{1}{M^2}\right)
$$

\subsection{Proof of Lemma \ref{asympt_lemma2}}
We first prove the statements for the multi-access scheme. 
We have $ M h_{min} c_{min} \leq \sum_{i=1}^M h_i c_i \leq M h_{max} c_{max}$
and 
$ M h_{min}^2 \sigma_{min}^2 \leq \sum_{i=1}^M h_i^2 \sigma_i^2  \leq M h_{max}^2 \sigma_{max}^2$.
Recall from Lemma \ref{lemma1} that $P_\infty$ is a decreasing function of $S= \bar{c}^2/\bar{r}$. 
If we choose $\alpha_i \in \{ +1, -1\}$ such that $\alpha_i c_i$ is positive for all $i$, we have
$$\frac{M h_{min}^2 \sigma_{min}^2+\sigma_n^2}{M^2 h_{max}^2 c_{max}^2} \leq \frac{\bar{r}}{\bar{c}^2} \leq \frac{M h_{max}^2 \sigma_{max}^2+\sigma_n^2}{M^2 h_{min}^2 c_{min}^2}$$
and by a similar calculation to (\ref{asymptotic_calc}) we can show that as $M \rightarrow \infty$
$$ \sigma_w^2 + \frac{a^2 h_{min}^2 \sigma_{min}^2}{h_{max}^2 c_{max}^2}\frac{1}{M} + O\left(\frac{1}{M^2}\right) \leq P_\infty \leq \sigma_w^2 + \frac{a^2 h_{max}^2 \sigma_{max}^2}{h_{min}^2 c_{min}^2}\frac{1}{M} + O\left(\frac{1}{M^2}\right)$$
If instead we choose $\alpha_i  \in \{1/\sqrt{M},-1/\sqrt{M}\}$ such that $\alpha_i c_i$ is positive for all $i$, we can similarly show that as $M \rightarrow \infty$
\begin{equation}
\label{asympt_bounds2}
 \sigma_w^2 + \frac{a^2 (h_{min}^2 \sigma_{min}^2+\sigma_n^2)}{h_{max}^2 c_{max}^2}\frac{1}{M} + O\left(\frac{1}{M^2}\right) \leq P_\infty \leq \sigma_w^2 + \frac{a^2 (h_{max}^2 \sigma_{max}^2+\sigma_n^2)}{h_{min}^2 c_{min}^2}\frac{1}{M} + O\left(\frac{1}{M^2}\right)
\end{equation}
In either case, as the upper and lower bounds both converge to $\sigma_w^2$ at a rate of $1/M$, $P_\infty$ itself will also do so. 

For the orthogonal scheme, a similar argument to the above shows that choosing $\alpha_i \in \{ +1, -1\}$ gives convergence of $P_\infty^o$ to $\sigma_w^2$ at the rate $1/M$ for general parameters.

To show that $P_\infty^o$ in general does not converge to a limit as $M \rightarrow \infty$, when using the scaling $1/\sqrt{M}$ in the orthogonal scheme, consider the following example. Suppose there are two distinct sets of ``symmetric'' parameters with behaviour as in (\ref{P_inf_orthogonal}), such that if all the sensors had the first set of parameters the error covariance would converge to $P_{\infty,1}^o$, and if all the sensors had the second set of parameters the error covariance would converge to $P_{\infty,2}^o$, with $P_{\infty,2}^o \neq P_{\infty,1}^o$. Then let the first $M_1$ sensors have the first set of parameters, the next $M_2$ (with $M_2 >> M_1$) sensors the second set, the next $M_3$ (with $ M_3 >> M_2$) sensors the first set, the next $M_4$ (with $ M_4 >> M_3$) sensors the second set, etc... Then $P_\infty^o$ would alternate between approaching $P_{\infty,1}^o$ and $P_{\infty,2}^o$, and will not converge to a limit as $M \rightarrow \infty$. 

\subsection{Proof of Lemma \ref{asympt_lemma3}}
With the multi-access scheme and the allocation (\ref{equal_power_alloc_1}), by defining 
$$ \alpha_{max}^2 = \frac{\gamma(1-a^2)}{c_{min}^2 \sigma_w^2 + \sigma_{min}^2 (1-a^2)}, \alpha_{min}^2 = \frac{\gamma(1-a^2)}{c_{max}^2 \sigma_w^2 + \sigma_{max}^2 (1-a^2)}$$
we can show similar to the proof of Lemma \ref{asympt_lemma2} that
$$\frac{M \alpha_{min}^2 h_{min}^2 \sigma_{min}^2+\sigma_n^2}{M^2 \alpha_{max}^2 h_{max}^2 c_{max}^2} \leq \frac{\bar{r}}{\bar{c}^2} \leq \frac{M \alpha_{max}^2 h_{max}^2 \sigma_{max}^2+\sigma_n^2}{M^2 \alpha_{min}^2 h_{min}^2 c_{min}^2}.$$
Hence as $M \rightarrow \infty$ we have
$$ \sigma_w^2 + \frac{a^2 \alpha_{min}^2 h_{min}^2 \sigma_{min}^2}{\alpha_{max}^2 h_{max}^2 c_{max}^2}\frac{1}{M} + O\left(\frac{1}{M^2}\right) \leq P_\infty \leq \sigma_w^2 + \frac{a^2 \alpha_{max}^2 h_{max}^2 \sigma_{max}^2}{\alpha_{min}^2 h_{min}^2 c_{min}^2}\frac{1}{M} + O\left(\frac{1}{M^2}\right).$$
The other cases can be treated similarly as in the proof of Lemma \ref{asympt_lemma2}. 

\end{appendix}


\bibliography{kalman_power_revision1}
\bibliographystyle{IEEEtran} 

\end{document}